\documentclass[fleqn,10pt]{wlscirep}
 \usepackage{amsmath}
 \usepackage{amssymb}
 \usepackage{color}
 \usepackage{amsthm}
 \usepackage{booktabs} 
 \usepackage{footnote}
 \usepackage{float}
 \usepackage{subfigure}
 \usepackage{graphicx}
 \usepackage{caption}
 \usepackage{soul}
 \usepackage{titlesec}
 \usepackage[ruled,commentsnumbered]{algorithm2e}  
 \usepackage{amsmath}  
 \usepackage{multirow}
\usepackage{threeparttable}
 
\newcommand{\xc}[1]{\mathbf{\dot{x}}^{(1)}(#1)}

\newtheorem{theorem}{Theorem}[section]  
\newtheorem{definition}{Definition}

\usepackage[utf8]{inputenc}
\usepackage[T1]{fontenc}
\title{ Novel Compositional Data's Grey Model for Structurally Forecasting Arctic Crude Oil Import}

\author[1]{Pan Qilong}
\author[1]{Yin Jieru}
\author[1,*]{Xiao Xinping}
\affil[1]{School of Science, Wuhan University of Technology, Wuhan 430070, China}

\affil[*]{xiaoxp@whut.edu.cn}

\begin{abstract}
The reserve of crude oil in Arctic area is abundant. Ice melting is making it possible to have intermediate access to the Arctic crude oil and its transportation. A novel compositional data's grey model is proposed in this paper to structurally forecast Arctic crude oil import. Firstly, the general accumulative operation sequence of multivariate compositional data is defined according to Aitchison geometry, then obtaining the novel model with the form of the compositional data vectors. Secondly, this paper studies the least square parameter estimation of the model.  The novel model is deduced and selected as the time-response expression of the solution. Thirdly, this papers infuses the novel model with traditional grey model to improve its robustness. Differential Evolution algorithm is introduced to determine the optimal value of the general matrix.  Lastly, two validation examples are provided for confirming the effectiveness of the novel model by comparing with other existing models, before being employed to forecast the crude oil import structure in China. The results show that the novel model provides better performance in all crude oil cases in short-term forecasting. Therefore, by using the new model, the China's development parameter is 0.5214 and Determination Factor of the novel model is 0.5999, which mean that the crude oil import structure of China is being changed. Specifically, the amount of crude oil imported from Arctic area are obviously increasing in the next 6 years, showing sufficient proof of the edge owned by Arctic area: abundant crude oil reserves and shortening transportation distance. 
\end{abstract}

\begin{document}

\flushbottom
\maketitle
%
%
\thispagestyle{empty}



\section{Introduction}
As the largest crude oil importer, China is facing the high cost of crude oil transportation from other countries. However, with the opening of Arctic routes, it may be economically meaningful to adjust the import proportions of various crude oil producing countries, optimize current transportation routes, and update vessel deployments to reduce the transportation cost and time of China's crude oil imports.
To achieve this goal, China needs to analyse its crude oil import structure and then make prediction for near-term changes. Therefore, the focus of this research is to develop a novel method for the crude oil import structure analysis and prediction.

\subsection{Literature Review}

Energy import prediction is a classical problem, and there are a series of advanced forecasting models about the energy system summarized in Table.\ref{introlist}, which shows that forecasting methodologies can be divided into three categories, namely, statistical analysis models, intelligent network models, and grey forecasting models. 
The common characteristic shared by all of the three types of models is the goal of predicting the real value trend of energy, that is, the accurate value of energies in the upcoming several years. Although the prediction on the white data can give a clear perspective to the trend of energy, the interaction among different regions nevertheless may be ignored.  Furthermore, a lack of behavior with overall analysis from global change is insufficient to reflect the structural shapes of a complete system and brings difficulties for a country to analyzing its energy import structure in a global perspective. An effective mathematical approach to the problem is compositional data, specializing in high dimensions and structural analysis \cite{Where}. Compositional data is initially popular in analyzing soil components and chemical elements, but hardly used in the energy structure analysis. To have an illustration of the special data, assuming we have already known that the sources of China's crude oil imports can be divided into five regions, then the contribution of the five areas can be expressed as [0.1829, 0.4734, 0.1193, 0.1128, 0.1115].
These datasets are characterized by multiple non-negative components, and the sum of the components is always 1, "the constant-sum constraint". In mathematics, the data describing the components is called compositional data, and its corresponding mathematical space is called "Simplex space".
However, in the Simplex space, the direct application of multivariate model to the compositional data will lead to incredible confusing results, such as negative compositions.

\begin{table}[htbp]
	\centering
	\caption{Review of hybrid energy forecasting implementations}
	\label{introlist}
	\small
	\begin{threeparttable}
	\resizebox{\textwidth}{!}{
		\begin{tabular}{lllllll}
			\toprule
			Literature & Study field    & Market          & Method                       & Scope  & Easy & Robust \\ \midrule
			Zhang et al.(2019)\cite{MA20191}    & Solar energy   & Japan           & Novel IGGM(1,1)        & Annual & Easy & High   \\
			Zhang et al.(2019)\cite{MA20192}     & Wind energy    & China           & Power-driven FAGM(1,1) & Annual & Easy & High   \\
			Wu et al.(2020)\cite{MA2020}     & Carbon dioxide & BRICS & CFNGM(1,1,k,c)         & Annual & Easy & High\\  
			Chen et al.(2020)\cite{chenhui}                        & Industrial energy & China  & Novel MCFGM(1,N) & Annual & Yes  & High   \\
			Wang et al.(2018)\cite{2018wang} & Electricity       & China  & Seasonal GM(1,1) & Annual & Yes  & High \\ 
			Xie et al.(2015)\cite{x2015}         & Total energy & China  & Grey Markov & Annual & Yes  & High   \\
			Pappas et al.(2008)\cite{pappas2008} & Electricity  & Greek  & ARIMA       & Daily  & Yes  & High   \\
			Azadah et al.(2011)\cite{Gas2011}    & Gas         & Middle East & ANFIS-SFA hybrid  & Annual & No  & Low    \\
			Huang et al.(2018)\cite{Huang2018}   & Crude Oil   & World       & Random wavelet NN & Daily  & Yes & Middle \\
			Kavaklioglu et al.(2009)\cite{k2009} & Electricity & Turkey      & Neural network    & Annual & No  & Low   \\
			Çevik et al.(2020)\cite{CEVIK2019122177} & Wind power & World & Artificial Neuro-Fuzzy Inference System & Annual & No & Middle \\
			Yan et al.(2014)\cite{2014Mid} & Electricity & PJM market & SVM-ARMAX     & Mid term & No   & Middle \\
			Cen et al.(2019)\cite{Cen2019}   & Crude oil price & West Texas     & Deep learning    & Long term & No   & Low    \\
			Zhao et al.(2020)\cite{Zhao2020} & Oil price       & China          & Time-varying     & Daily     & Yes  & Middle \\
			Lin et al.(2019)\cite{LIN2020123532} & Crude oil price & China & novel hybrid long memory GARCH-M & Annual & Yes & Middle\\
				\bottomrule
    	\end{tabular}}
		\begin{tablenotes} \scriptsize
			\item[1] {'Easy' means the method is easy to realize without calling the existing toolboxes or code packages.}
			\item[2] {'Robust' usually means that the results of forecasting method hardly change even with the small-scale disturbance both in the data and parameters.}
		\end{tablenotes}
\end{threeparttable}
\end{table}


At present, two ways of avoiding the problem are proposed, that is, transformation and Aitchison geometry.
As for transformations, one of the most effective is isometric logarithmic ratio \emph{ilr} transformation, which can not only solve the indeterminate choice of the denominator; but also circumvent the linearly correlated data after transformation. Not only does \emph{ilr} transformation eliminate the fixed-sum constraints and redundant dimensions of the compositional data, but also constructs an orthogonal transformation from a monomorphic space to an Euclidean space. The transformation  ensures the addition and multiplication of vectors, meanwhile, and maintains the good properties of the inner product, the modulus, the angle and the point distance of the vectors. To that end, the equivalent transformation from Simplex space $S^D$ to Euclidean space $R^{D-1}$ is thus realized, ensuring the rationality of modeling in the transformed Euclidean space.
As for the Aitchison geometry, Aitchison proposed the definition of perturbation operation in the Simplex space \cite{roleofperturbation}, and then the inner product, distance, and relations of three logarithmic ratio transformations in the Simplex space were proposed successively \cite{Where}. Simplex space proved to be a complete Hilbert space, and its algebraic system was then Aitchison geometry. 
For compositional data's modeling, Wang et al. modeled the  linear regression relationship for compositional data using Aitchison geometry. \cite{multiregression};
Dumuid et al. studied the impact of exercise duration, sedentary duration, and sleep duration on human health in the Simplex space \cite{dumuid2018compositional}. 
Pawlowsky summarized the spatial statistical analysis and the use of the Kriging interpolation method in the Simplex space.\cite{pawlowsky2016spatial}
This Aitchison geometry contributes to saving the transformation steps and ensuring that the data in the modeling process consistently meet the fixed-sum constraints.

\subsection{Research Gap}

After introducing the compostional data into prediction of energy import structure, the proper consideration of various models is taken. 
As Table.\ref{introlist} shows, three main models are in short list ( statistical analysis models, intelligent network models, and grey forecasting models).  Regression models, such as Autoregressive Integrated Moving Average (ARIMA), have gained popularity in medium- and long-term energy consumption prediction.\cite{pappas2008,XU201742} However, a prominent limitation is that their forecasting performance is highly dependent on sufficient samples and variables.\cite{LENG2020124663} Furthermore, statistical regression models may lose its prediction efficacy when the sample data cannot satisfy statistical assumptions \cite{pappas2008,2020xqz}. Additionally, other scholars have begun to focus their attention on Artificial intelligence (AI) approaches \cite{DOSS2020122656}, which involve Neural network, Deep learning method. The models demonstrate high flexibility in forecasting, estimating, and overcoming noisy data. Then again, AI methods are difficult to explain the economic significance of coefficients and its accuracy heavily relies on the number of the training sample data \cite{Gas2011}. Yet, the source data sequence of effective observations concerning certain energies, such as the crude oil import, is relatively limited in size and does not always conform to a certain statistical distribution. Besides, the compositional data is a type of 'grey' data due to its uncertainty meaning of real value. Statistical regression models and AI models are, therefore, not appropriate prediction tools for energy consumption forecasting in certain circumstance.\cite{ZHU2019392} The grey system theory has proved its efficiency as an estimation method under such conditions for forecasting annual energy consumption.\cite{2020xqz, Qin2020A}

Grey forecasting theory was proposed by Professor Deng in
1982 to deal with uncertain problems containing incomplete information.\cite{TANG2020123615} Among the theory, discrete grey model gradually becomes an appropriate prediction method to settle clean energy consumption forecast issues\cite{xx1,xx2,xx3}. For instance, Yao conducted research on crude oil price prediction based on LSTM network and GM (1,1) model. \cite{GM1}
Ding estimated Chinese energy-related CO2 emissions through employing a novel discrete grey multi-variable prediction model.\cite{d2020}
Wang proposed two new combined MNGM-ARIMA and MNGM-BPNN models on the basis of 'error correction + quadratic modeling' strategy to model carbon emission trajectory of China, US and India.\cite{wang20202} Notably, the gaps to previous scholars research are listed as bellow, followed by their solutions.

\begin{enumerate}
	\item Few people had a study on structural analysis with grey system theory under compositional data, especially through Aitchison geometry. That being said, creating novel grey models on compositional data is worth trying. 
	\item The way of directly transforming compositional data in Simplex space into Euclidean space is easily understood. Although the GM(1,1) model based on \emph{ilr} transformation ensures the heterogeneity of every composition, it nevertheless ignores the correlation of different dimensions, such as the independent control coefficient for all components. The use of Aitchison geometry may be a proper approach to improve the traditional models.
	\item If using Aitchison geometry, which allows for homogeneity of every composition and yet contains only add and subtract, scalar multiplication and inner product, it gives no way to define  GM(1,1)'s continuous form as it does in Euclidean space. The discrete form of GM(1,1) model can be considered.
\end{enumerate} 

\subsection{Main Work}

Overall, considering the high dimension and limited sample of China's crude oil data due to only a few years after opening of Arctic passage and grey features of compositional data, this paper proposed the discrete GM(1,1) model under compositional data to have prediction for import structure of crude oil in China. The main work of this paper lies in the following aspects:

Firstly, the modeling mechanisms for crude oil import structure 
forecasting according to overall change rate and heterogeneity between regions are considered to establish a new forecasting and analyzing model.

Secondly, the discrete form of GM(1,1) is adopted and extended to the Simplex space with Aitchison geometry. Moreover, generalized accumulation matrix in the GM(1,1) model is solved by Differential Evolutionary algorithm (DE algorithm) and proves to enhance models' robustness over different data sets. 

Lastly, the verification and efficacy of the new model is testified by two validation examples, and then the model is applied to forecast the crude oil import structure of the China in the next six years. The results are used to substantiate the crude oil edge owned by Arctic area: abundant crude oil reserves and shortening transportation distance.

The outline of this paper is organized as follows. Section 2 introduces the methodology including the formulation of
the proposed model and its solution algorithm. In sections 3,
two datasets are implemented to validate the proposed method and another dataset is used for China's crude oil application. Finally, the paper is concluded in section 4. The concepts, abbreviations and their corresponding definitions are listed in Table.\ref{abbrev}.
\begin{table}[!htb]
\centering
\caption{Concepts/abbreviations and corresponding definitions}
\label{abbrev}
\scriptsize
\resizebox{\textwidth}{!}{
\begin{tabular}{ll}
\toprule
Concepts/Abbreviations & Definition                    \\ \hline
GADGMSS            & General-Accumulation Discrete GM(1,1) model in Simplex Space            \\
TGMI                   & Traditional GM(1,1) model with \emph{ilr}-transformation   \\
IGADGM                 & Infused General-Accumulation Discrete GM(1,1) model \\
DE algorithm                    & Differential Evolution algorithm                    \\
Compositional data & A data type, characteristics of proportions and constant-sum constrict  \\
Simplex Space          & Mathematical space of compositional data            \\
Aitchison geometry     & Geometry in Simplex space                           \\
\emph{ilr} transformation & isometric logratio transforming compositional data from Simplex space to Euclidean space \\
Development coefficient  & Overall change rate year-on-year for a complete system \\
\emph{DF} & Determination Factor for weighing the proportion of the results from the two models\\

\bottomrule
\end{tabular}}
\end{table}

\section{Methodology} \label{sec2}

In this section, we primarily build General-Accumulation Discrete GM(1,1) Model with Aitchison Geometry (GADGMSS) in Simplex space, and prove its consistent meaning of parameters solution in Euclidean space. Meanwhile, considering homogeneity of the model with Aitchison geometry, we propose to infuse GADGMSS with Traditional GM(1,1) model with ilr (TGMI) based on their error weight, that is, the Infused General-Accumulation GM(1,1) model (IGADGM), which improves robustness of both GADGMSS and TGMI. Before the modeling, some necessary preliminaries are needed to be prepared.
		
\subsection{Preliminaries to Simplex Space}

In order to make the notation more clear, this paper adopts a notation system, where [...],($\cdot$,$\cdot$) indicate compositional data and their inner product in Simplex space, and (...),  $\langle \cdot$,$\cdot \rangle$ indicate the same operation in real space $\mathbb{R}$. 
  	
\subsubsection{Aitchison Geometry}
	
	The \textit{D}-part unit Simplex space is defined as follows:
	\begin{equation}
	S^D=\{ [x_1,\dots,x_D]:x_i>0 (i=1,2,\dots,D),x_1+\dots+x_D =1 \}
	\end{equation} 
	
	Basic operations in the Aitchison Geometry and their properties are summarized in Ref.\cite{Where,roleofperturbation}, and there are necessary ones listed. Given any two \textit{D}-part compositions $\mathbf{x,y}\in S^D$, their perturbation is 
	\begin{equation}\label{oplus}
		\mathbf{x \oplus y} = \mathcal{C} (x_1y_1,x_2y_2,\dots,x_Dy_D),
	\end{equation}
	where $\mathcal{C}$ means closure or normalizing operation defined as follows in which the elements of a positive vector are divided by their sum,
	\begin{equation}
	\mathcal{C}(x_1,x_2,\dots,x_D)=(\frac{x_1}{\sum_{i=1}^{n}x_i},\frac{x_2}{\sum_{i=1}^{n}x_i},\dots,\frac{x_D}{\sum_{i=1}^{n}x_i}).
	\end{equation}
	
	Given a real number $a \in \mathbb{R}$, the power transformation is 
	\begin{equation}\label{otimes}
		a\otimes \mathbf{x}= \mathcal{C}(x_1^a,x_2^a,\dots,x_D^a).
	\end{equation}
 	Note that $\otimes$ and $\oplus$ are defined for the analogy with the familiar operation of translation and scalar multiplication of vectors in the real space $\mathbb{R}^D$. Apparently, the inverse perturbation $\ominus$ is defined as 
	\begin{equation}
	       \mathbf{x\ominus y} = \mathcal{C} [\frac{x_1}{y_1},\frac{x_2}{y_2},\dots,\frac{x_D}{y_D}].
	\end{equation}
	
	The  inner product $(\mathbf{x,y})_S$ is defined by
	\begin{equation}
		(\mathbf{x,y})_S=\sum_{i=1}^{D}log\frac{x_i}{g(\mathbf{x})}log\frac{y_i}{g(\mathbf{y})}
	\end{equation}
	where $g(\cdot) = \sqrt[D]{x_1x_2...x_D}$ is the geometric mean of all parts of $\mathbf{x}$. An important role is the relation between any two of norm, inner product, and distance \cite{multiregression}, which will be crucial to the understanding of the later proof.
	\begin{gather}\label{innerproduct1}
		||\mathbf{x}||^2_S = (\mathbf{x,x})_S; \quad
		||\mathbf{x\ominus y}||_S^2 = \Delta_S^2(\mathbf{x,y})
	\end{gather}

\subsubsection{ilr Transformations}

	The direct application of multivariate model to $D$-part compositional data will lead to incredible confusing results, such as negative compositions \cite{Where}. In view of the importance of compositional data in analytical prediction \cite{log_important1,log_important2}, several of ways are proposed, for instance, the family of logratio transformation, to eliminate the constraint of compositional data before applying the data into Grey model. These transformation methods mainly include \emph{alr} transformation\cite{compositionaldistance}, \emph{clr} transformation. Importantly, \emph{ilr}\cite{PARAMETRIC}  is a more effective method to overcome the closure effect of compositional data.\cite{LT}. 
	For $ \mathbf{x} \in S^D$, the transformation $\textit{ilr}(\cdot)$ can be concisely written in matrix form,
	\begin{equation}
	\mathbf{y}=ilr(\mathbf{x})= \mathbf{H'}\mathbf{x}
	\end{equation}
	where $\mathbf{H}= (\mathbf{e_1,e_2,...,e_{D-1}})$ is a $D\times(D-1)$ orthonormal basis matrix, with
	\begin{equation}\label{ilrtrans}
	\mathbf{e_i} = \mathcal{C} (exp(\underbrace{\sqrt{\frac{i}{i+1}},\dots,\sqrt{\frac{i}{i+1}}}_i,-\sqrt{\frac{i}{i+1}},0,\dots,0).
	\end{equation}
 
 	The Equation.\ref{ilrtrans} indicates an desired property $ilr(\mathbf{e_i})=\vec{e_i}$ ($i=1,2,\dots,D-1$), where $\mathbf{e_i}$ is the basis in Simplex space and $\vec{e_i}$ is the canonical basis in $\mathbb{R}^{D-1}$.  Moreover, since the length of vectors and the angle of between them are defined through inner product, orthonormal transformations preserve these qualities.
 	
 	The inverse \textit{ilr} transformation corresponds to the expression of $\mathbf{x}$ in the reference basis of $S^D$:
 	\begin{equation}
 		\mathbf{x}= \textit{ilr}^{-1}(y)=\bigoplus_{i=1}^{D-1}(\langle \mathbf{y},\vec{e_i}\rangle\otimes \mathbf{e_i}),
 	\end{equation}	  
	where the symbol $\bigoplus$ stands for the repeated perturbation, as well as $\langle\mathbf{y},\vec{e_i}\rangle=(\mathbf{x},\mathbf{e_i})=y_i$, $i=1,2,...,D-1$. \cite{ilr1,ilr2}

\subsubsection{Compositional Data Notations System}
	According to Wang et al.,\cite{multiregression} this article chooses to use their vector system as to formulate new models. A D-dimensional compositional data vector with n sample points is identified with,
	\begin{equation}
		\mathbf{X}^{(n)}=
		\begin{pmatrix}
		\mathbf{x}^{(n)}_1 \\
		\vdots \\
		\mathbf{x}^{(n)}_n
		\end{pmatrix}=
		\begin{pmatrix}
		\mathbf{r}^{(n)}_1 & \dots &\mathbf{r}^{(n)}_D
		\end{pmatrix}=
		\begin{pmatrix}
		x^{(n)}_{11} & \dots & x^{(n)}_{1D}\\
		\vdots & \ddots & \vdots\\
		x^{(n)}_{n1} &  \dots  &  x^{(n)}_{nD}
		\end{pmatrix}	
	\end{equation}
	\noindent where n means the times of accumulation from original data, e.g., $\mathbf{X}^{(0)}$ is original data and $\mathbf{X}^{(1)}$ is accumulated data with one time. Notably, $\mathbf{x}^{(n)}_i(1 \leq i \leq n)$ can be visually treated as one sample point vector with the quality of $\sum_{j=1}^{j=D}x^{(n)}_{ij}=1$; and similarly, $\mathbf{r}^{(n)}_j(1 \leq j \leq D)$ as a variable vector containing values from one dimension of all points, as well as $ x^{(n)}_{ij} \geq 0 $.  
	
 $\forall \mathbf{X,Y} \in S_n^D, \beta \in \mathbb{R}$, the corresponding matrix operations are,
	\begin{gather}
		\mathbf{X}\oplus\mathbf{Y}=
			\begin{pmatrix}
				\mathbf{x}_1 \oplus \mathbf{y}_1,
				\mathbf{x}_2 \oplus \mathbf{y}_2,
				\dots,
				\mathbf{x}_n \oplus \mathbf{y}_n
			\end{pmatrix}^T \\
		\beta\otimes\mathbf{X}=
		\begin{pmatrix}
			\beta \otimes \mathbf{x}_1,
			\beta \otimes \mathbf{x}_2,
			\dots,
			\beta \otimes  \mathbf{x}_n
		\end{pmatrix}^T \\
		(\mathbf{X},\mathbf{Y})_S = 
		\sum_{i=1}^{n}(\mathbf{x}_i,\mathbf{y}_i)_S; \quad
		||\mathbf{X}||^2_S=(\mathbf{X},\mathbf{X})_S; \quad		\Delta_S^2(\mathbf{X},\mathbf{Y})=||\mathbf{X}\ominus\mathbf{Y}||^2_S. \label{innerproduct}
	\end{gather}	
	Besides these basic operations in Aitchison geometry, we extend the power operation into matrix multiplication as follows.
	\begin{definition}
		$\forall \mathbf{X} \in S_n^D, \mathbf{B}=(\beta_{ij})_{n\times n}\in \mathbb{R}^{n}$, the power transformation $\otimes$ is defined as,
		\begin{equation}
			\mathbf{B}\otimes\mathbf{X}=
			\begin{pmatrix}
				\bigoplus_{j=1}^{n}\beta_{1j}\otimes \mathbf{x}_j\\
				\bigoplus_{j=2}^{n}\beta_{2j}\otimes \mathbf{x}_j\\
				\vdots\\
				\bigoplus_{j=n}^{n}\beta_{nj}\otimes \mathbf{x}_j\\
			\end{pmatrix}
		\end{equation}
	where $\beta_{ij}\otimes \mathbf{x}_j, i,j=1,2,...,n$ satisfies Equation.\ref{oplus}. In grey theory, this process is called General-Accumulation technique, making data more regular before modeling.
	\end{definition}

\subsubsection{General-Accumulation Discrete GM(1,1) Model} \label{GM(1,1)}

 The traditional accumulation is substituted  with General-Accumulation in following way. let $\mathbf{B}\in \mathbb{R}^n$,   
\begin{equation}\label{Blimits}
\mathbf{B}=\begin{pmatrix}
b_{11} & 0 & \dots & 0 & 0\\
b_{21} & b_{22} & \dots  & 0 & 0\\
\vdots & \ddots & \ddots& \ddots & \vdots\\
b_{n-1,1} & b_{n-1,2} & \dots  & b_{n-1,n-1} & 0\\
b_{n1} & b_{n2} & \dots  & b_{n,n-1} & b_{nn}
\end{pmatrix}
\end{equation}
where the determinant $||\mathbf{B}||\ne 0$ ensures the existence of inverse of the General-Accumulation matrix.
\begin{equation} \label{preaccumulation}
x^{(1)}{(k)} =\sum_{i=1}^{k}{\beta_{ik}x^{(0)}{(i)}},
\end{equation}
can be written in matrix form: $X^{(1)}=\mathbf{B}X^{(0)}$, where $X^{(1)}$ = $(x^{(1)}(1)$ $,x^{(1)}(2)$ $,...$ $,x^{(1)}(n))^T,$ $i=1,2$.

The Discrete GM(1,1) model concerns itself with the following way \cite{theorem2}: 
\begin{equation}
x^{(1)}(k+1)=\beta_1 x^{(1)}(k)+ \beta_2
\end{equation}\label{greymodel}
\noindent where $x^{(1)}{(k)}$ and $x^{(1)}(k+1)$ $(k=1,2,...,n-1)$, are the accumulated data.

The solution of the Discrete GM(1,1) model is,
\begin{equation}
	\hat{x}^{(1)}(k+1)=\beta^k_1x^{(0)}(1)+\frac{1-\beta^k_1}{1-\beta_1}\beta_2, \qquad k=1,2,...,n-1
\end{equation}

The inverse of decrement here is
\begin{equation}
\hat{X}^{(0)}=\mathbf{B}^{-1}\hat{X}^{(1)}.
\end{equation}

\subsection{Forecasting Methods}
The high dimension of compositional data makes it impossible to achieve the analytical solution by solving the approximate time-response formula. That being said, the vector cannot now be differentiated directly. Therefore, the solution is to introduce Discrete GM(1,1) model.

\subsubsection{GADGMSS Modeling}

Different from the traditional grey models, the General-Accumulation in the Simplex Space is defined as,
\begin{gather} 
\mathbf{X}^{(1)}=\mathbf{B}\otimes\mathbf{X}^{(0)},\\ 
\mathbf{\hat{X}}^{(0)}=\mathbf{B}^{-1}\otimes\mathbf{\hat{X}}^{(1)}, \label{inversecomposition}
\end{gather}

GADGMSS is presented in form of discretion in following way:
\begin{equation}
	\mathbf{x}^{(1)}(k+1)=\beta_1 \otimes \mathbf{x}^{(1)}(k) \oplus \beta_2 \otimes \mathbf{I} 
\end{equation}
\noindent where $\mathbf{I}$ means a unit vector, with every value equal to 1. That said, frustratingly, there is no such a counterpart of $\mathbf{I}$ in Simplex space. The constant, $\beta_2$, needs to be eliminated. This paper, therefore, applies the centralization of compositional data \cite{chenhui,multiregression} to solve the problem.

\begin{definition} \label{center}
	$\forall \mathbf{X}^{(1)} \in S_n^D$, the compositional center $\overline{\mathbf{r}}^{(1)}$ is defined as follows:
\begin{equation}
\overline{\mathbf{r}}^{(1)}=\mathcal{C}
	\begin{pmatrix}
	 	\overline{r}^{(1)}_1 & \overline{r}^{(1)}_2 & \dots & \overline{r}^{(1)}_D
	\end{pmatrix}
\end{equation}

\noindent where $\overline{r}^{(1)}_j=g(\mathbf{r}^{(1)}_j) = \sqrt[n]{x^{(1)}_{1j} x^{(1)}_{2j} \dots x^{(1)}_{nj}} ,1 \leq j \leq D$, and $\mathcal{C}$ means closure operation. The compositional center of the sample is abbreviated as $ \mathbf{E} (\mathbf{X}^{(1)})$. 
\end{definition}
\begin{definition} \label{centralization} Based on its center, compositional centralization is concluded: 
\begin{equation}
	 \mathbf{\dot{X}}^{(1)}= {\it{Cen}}
	 (\mathbf{X}^{(1)})=\mathbf{X}^{(1)} \ominus 
	 \begin{pmatrix}
	 		\mathbf{\overline{r}}^{(1)} \\
	 		\vdots \\
	 		\mathbf{\overline{r}}^{(1)}
	 \end{pmatrix}
\end{equation}

\noindent which is equally expressed as $\mathbf{\dot{x}}^{(1)}_i  = \mathbf{x}^{(1)}_i \ominus \overline{\mathbf{r}}^{(1)}  ,  (1 \leq i \leq n)$, abbreviated as ${\it{Cen}}
(\mathbf{X}^{(1)})$.
\end{definition}  

Through the application of the compositional centralization on $\mathbf{X}^{(1)}$, the constant $\beta_1$ is eliminated, which is formally analogous to that of multivariate regression model in Simplex Space\cite{multiregression,spatial}. The Discrete GM(1,1) model is formulated as follows:
\begin{equation} \label{equa1}
     \mathbf{\dot{x}}^{(1)}(k+1)=\beta_1  \otimes \mathbf{\dot{x}}^{(1)}(k), k=1,2,...n-1  
\end{equation}

	The Ordinary Least Squares method (OLS) is used to solve the parameter $\beta_1$ as the traditional GM(1,1) does. The non-negative, symmetry, and linearity in Aitchison geometry have been proven \cite{multiregression}.

\begin{theorem}\label{coef}
	Assuming 
	\begin{gather}
		\mathbf{\dot{X}}^{(1)}_1=
		\begin{pmatrix}
			\mathbf{\dot{x}}^{(1)}(2)\\
			\mathbf{\dot{x}}^{(1)}(3)\\
			\vdots \\
			\mathbf{\dot{x}}^{(1)}(n)
		\end{pmatrix},
		\mathbf{\dot{X}}^{(1)}_0=
		\begin{pmatrix}
			\mathbf{\dot{x}}^{(1)}(1)\\
			\mathbf{\dot{x}}^{(1)}(2)\\
			\vdots \\
			\mathbf{\dot{x}}^{(1)}(n-1)		
		\end{pmatrix},
	\end{gather}
	then, the estimate $\hat{P}=(\hat{\beta}_1)^T$ of GADGMSS satisfies
	\begin{equation}
		\hat{P}=(\hat{\beta}_1)^T=(\mathbf{\dot{X}}^{(1)}_0,\mathbf{\dot{X}}^{(1)}_0)^{-1}_S(	\mathbf{\dot{X}}^{(1)}_0,	\mathbf{\dot{X}}^{(1)}_1)_S
	\end{equation}
\end{theorem}
	$Note:$ From the view of mathematical expression, the very essence between the estimate of $\hat{P}=(\mathbf{\dot{X}}^{(1)}_0,\mathbf{\dot{X}}^{(1)}_0)^{-1}_S(	\mathbf{\dot{X}}^{(1)}_0,	\mathbf{\dot{X}}^{(1)}_1)_S$ in GADGMSS and that in GM(1,1) (\textit{for example:} $\hat{P}=(B^TB)^{-1}B^TY$), both of which are substantially equivalent, is to be clear with following proof.
	
	\noindent $\mathbf{Proof.}$
	The GADGMSS is rewritten in matrix form as 
\begin{gather}
	\begin{pmatrix}
		\mathbf{\dot{x}}^{(1)}(2)\\
		\mathbf{\dot{x}}^{(1)}(3)\\
		\vdots \\
		\mathbf{\dot{x}}^{(1)}(n)
	\end{pmatrix}
	=
	\begin{pmatrix}
	\mathbf{\dot{x}}^{(1)}(1)\\
	\mathbf{\dot{x}}^{(1)}(2)\\
	\vdots \\
	\mathbf{\dot{x}}^{(1)}(n-1)		
	\end{pmatrix}
	\otimes (\hat{\beta}_1),
\end{gather}
that is,
\begin{equation}
		\mathbf{\dot{X}}^{(1)}_1 = 	\mathbf{\dot{X}}^{(1)}_0 \otimes P
\end{equation}

In the above equation set, $\mathbf{\dot{X}}^{(1)}_1, \mathbf{\dot{X}}^{(1)}_0$ are known, and $P$ is the undetermined parameter. Moreover, the set includes n-1 equations with only one variable and the constraint $n>1$. Coupled with these facts, the set has no solution when incompatible, while OLS can be used to estimate $P$. 

With $\beta_1 \otimes \mathbf{\dot{x}}^{(1)}(k+1)$ substituting for $\mathbf{\dot{x}}^{(1)}(k), k=1,2,...,n-1$, the sequence of error $\boldsymbol{\epsilon}$ is obtained, 
\begin{equation}
	\boldsymbol{\epsilon} = 	\mathbf{\dot{x}}^{(1)}(k+1) \ominus \beta_1 \otimes \mathbf{\dot{x}}^{(1)}(k)
\end{equation}
\noindent where $\boldsymbol{\epsilon}$ is the sampling error between two adjacent compositions. Furthermore, minimum of square errors ($SSE$) is drawn as below: 
\begin{equation} \label{SSE}
min(SSE) =min {\left\| \mathbf{\dot{X}}^{(1)}_1 \ominus \hat{\beta}_1 \otimes \mathbf{\dot{X}}^{(1)}_0  \right \|_S^2 }  
\end{equation}

With the properties in Equation.\ref{innerproduct1}, it is easy to rewrite the expression of $SSE$ in matrix form, where matrix derivative can literally work. 
\begin{equation}
\begin{aligned}
   SSE =&(\mathbf{\dot{X}}^{(1)}_1\ominus  \mathbf{\dot{X}}^{(1)}_0 \otimes P,\mathbf{\dot{X}}^{(1)}_1\ominus \mathbf{\dot{X}}^{(1)}_0\otimes P)_S  \\
	= & (\mathbf{\dot{X}}^{(1)}_1,\mathbf{\dot{X}}^{(1)}_1)_S-2(\mathbf{\dot{X}}^{(1)}_1\otimes P,\mathbf{\dot{X}}^{(1)}_0)_S + P^T\otimes(\mathbf{\dot{X}}^{(1)}_0,\mathbf{\dot{X}}^{(1)}_0)_S\otimes P   \label{see}
\end{aligned}
\end{equation}
The values of $\hat{P}$ is eventually obtained with Matrix Derivation,
\begin{equation}
\hat{P}=(\mathbf{\dot{X}}^{(1)}_0,\mathbf{\dot{X}}^{(1)}_0)^{-1}_S(	\mathbf{\dot{X}}^{(1)}_0,	\mathbf{\dot{X}}^{(1)}_1)_S. 
\end{equation} $\Box$

$\hat{P} = (\hat{\beta}_1)^T$ can as well be written in the explicit expression as follows with matrix expansion, 
\begin{equation}
\hat{\beta}_1=\frac{\sum_{k=1}^{n-1}(\xc{k+1},\xc{k})_S}{\sum_{k=1}^{n-1}(\xc{k},\xc{k})_S}.
\end{equation}

Note that despite only one variable in the $P=(\beta_1)^T$, Theorem. \ref{coef} is not intentionally to cast a mist on its conclusion with matrix form and instead serves as an useful tool when dealing with multivariate Grey Model. To give an illustration, Theorem. \ref{coef} works on the condition of $(\mathbf{X}_0,\mathbf{X}_1,...,\mathbf{X}_n)\otimes P$ with $P=(\beta_0,\beta_1,...,\beta_n)^T$, which will be discussed in the future research  providing that the article's key still lies on the GADGMSS after all. 

With the estimate of $\hat{P}$, the results of predictive value $\mathbf{\hat{\dot{X}}}^{(1)}_1$ are shown as,
\begin{equation}
	\mathbf{\hat{\dot{X}}}^{(1)}_1 = \mathbf{\dot{X}}^{(1)}_0\otimes P,
\end{equation}
which can be written in explicit expression as $\mathbf{\hat{\dot{x}}}^{(1)}(k+1)=\hat{\beta}_1\otimes\mathbf{\dot{x}}^{(1)}(k)(k=1,2,...,n-1)$. The data is still in type of centralization and accumulation, which cannot be directly compared with the original data in terms of error evaluation. Decentralization and decrement are then implemented on these outcomes. 
\begin{definition}
	$\forall \mathbf{\hat{\dot{X}}}^{(1)}(k) \in S_n^{D}$, the decentralization of compositional vectors is defined as :
	\begin{equation}
	\mathbf{\hat{X}}^{(1)}={\it{DeCen}}
	(\mathbf{\hat{\dot{X}}}^{(1)})= \mathbf{\hat{\dot{X}}}^{(1)} \oplus 
	\begin{pmatrix}
	\mathbf{\overline{r}^{(1)}} \\
	\vdots \\
	\mathbf{\overline{r}^{(1)}}
	\end{pmatrix}
	\end{equation}
\end{definition}
\noindent which is also expressed as $\mathbf{\mathbf{\hat{x}}^{(1)}}(k)  = \mathbf{\hat{\dot{x}}}^{(1)}(k) \oplus \overline{\mathbf{r}} ,  ( k=1,2,...,n-1)$. Note that $ \mathbf{\overline{r}}$ is compositional center, which has been illustrated in Definition.\ref{center}.

Eventually, from the Equation.\ref{inversecomposition} the predicted $\mathbf{\hat{X}}^{(0)}$ are obtained: 
\begin{equation}\label{outcomes}
			\mathbf{\hat{X}}^{(0)}=\mathbf{B}^{-1}\otimes\mathbf{\hat{X}}^{(1)}
\end{equation}

\subsubsection{Parameter Explanations}
	It is worth considering the meaning of the estimated parameter $\hat{\beta_1}$, which may give a vivid explanation of how the $\hat{\beta}_1$ is related with real values instead of the compositions. Considering the operation rules of Aitchison geometry in Simplex Space, Theorem.\ref{coef}'s conclusion can be written as,
	\begin{equation}\label{theorem2}
		\hat{\dot{x}}^{(1)}_j(k+1) = \frac{{\dot{x}^{(1)}_j(k)}^{\hat{\beta}_1}}{\sum_{j=1}^{D}{\dot{x}^{(1)}_j(k)}^{\hat{\beta}_1}}
	\end{equation}
	where $	\hat{\dot{x}}^{(1)}_j(k+1)$ and ${\dot{x}^{(1)}_j(k)}$ are, respectively, the components of $	\mathbf{\hat{\dot{X}}}^{(1)}_1$ and $\mathbf{\dot{X}}^{(1)}_1$, $k=1,2,...,n$ and $j=1,2,..,D$. These data only represent the proportion of the population rather than the real values of the population. Therefore, we suppose $\mathbf{G}_i=(G_{i1},...,G_{iD})$ be the real value, and $\mathbf{v}_i=\mathcal{C}(G_{i1},...,G_{iD})$ be the $\mathbf{G}_i$'s compositions, i.e., $v_{ij}=G_{ij}/\sum_{j=1}^{D}G_{ij}$. For instance, assuming $v_{i1},v_{i2},v_{i3}$ denote the components in percentage of the amount of crude oil in Arctic, America, and Middle East, respectively. $\mathbf{G}_{i1},\mathbf{G}_{i2},\mathbf{G}_{i3}$ can then be regarded as the real amount of the crude oil in these areas, respectively.

	\begin{theorem}
		Assuming that $\mathbf{G}_i=(G_{i1},...,G_{iD})$ is a given vector sequence satisfying $\mathbf{v}_i=\mathcal{C}(G_{i1},...,G_{iD})$; $\mathbf{\hat{G}}_i=(\hat{G}_{i1},...,\hat{G}_{iD})$ is a predicted vector sequence satisfying $\mathbf{\hat{v}}_i=\mathcal{C}(\hat{G}_{i1},...,\hat{G}_{iD})$, then, 
		\begin{equation}
			\hat{\beta}_1=\frac{\Delta log(G_{(k+1)j})/G_{(k+1)j'})}{\Delta log(G_{kj})/G_{kj'})}
		\end{equation}
		where $\Delta(\cdot)$ represents the change of the variable, and $\mathcal{C}(\cdot)$  means the closure operation.
	\end{theorem}
	\noindent $\mathbf{Proof.}$ From the Equation.\ref{theorem2}, it can be deduced that,
	\begin{equation}
		\frac{\hat{G}_{(k+1)j}}{\sum_{j=1}^{D} \hat{G}_{(k+1)j}} = \frac{{G_{kj}}^{\hat{\beta}_1}}{\sum_{j=1}^{D}{G_{kj}}^{\hat{\beta}_1}}
	\end{equation}
	To take logarithm on both sides of the above equation,
	\begin{gather}\label{proof2.2.1}
		log(\hat{G}_{(k+1)j})-log(\sum_{j=1}^{D}\hat{G}_{(k+1)j})
		=
		{\hat{\beta}_1}log({G_{kj}})-log(\sum_{j=1}^{D}{G_{kj}}^{\hat{\beta}_1}) \\ \label{proof2.2.2}
	log(\hat{G}_{(k+1)j'})-log(\sum_{j=1}^{D}\hat{G}_{(k+1)j})
	=
	{\hat{\beta}_1}log({G_{kj'}})-log(\sum_{j=1}^{D}{G_{kj}}^{\hat{\beta}_1})
	\end{gather}
	where $k,j,j'$ are satisfied the condition that $k=1,2,...,n$ means the number of sample; and $j\ne j',j,j'=1,2,...,D$ means the dimensions of compositional data.  To subtract Equation.\ref{proof2.2.2} from the Equation.\ref{proof2.2.1},
	\begin{equation}
		log(\hat{G}_{(k+1)j}/\hat{G}_{(k+1)j'})=
		{\hat{\beta}_1}log({G}_{kj}/{G}_{kj'})
	\end{equation}
	Then, it can obtain that,
	\begin{equation}
		{\hat{\beta}_1}=\frac{\Delta log(\hat{G}_{(k+1)j}/\hat{G}_{(k+1)j'})}{\Delta log({G}_{kj}/{G}_{kj'})}
	\end{equation}
	$\Box$
	
	Note that $\Delta log(\hat{G}_{(k+1)j}/\hat{G}_{(k+1)j'})$ $
	\approx$ $ 
	\Delta (\hat{G}_{(k+1)j}/\hat{G}_{(k+1)j'})$ 
	and 
	$ \Delta log({G}_{kj}/{G}_{kj'}) $ $\approx$ $ \Delta ({G}_{kj}/{G}_{kj'})/({G}_{kj}/{G}_{kj'})$ 
	when 
	$\hat{G}_{(k+1)j}/\hat{G}_{(k+1)j'}$ and ${G}_{kj}/{G}_{kj'}$ tend to 1. 
	Consequently, $\Delta log(\hat{G}_{(k+1)j}/\hat{G}_{(k+1)j'})$ and $\Delta log({G}_{kj}/{G}_{kj'})$ can be regarded as the change rate of $\hat{G}_{(k+1)j}/\hat{G}_{(k+1)j'}$ 
	and ${G}_{kj}/{G}_{kj'}$, respectively. In this sense, $\hat{\beta}_1$ implies the notion of overall change rate year-on-year; that is, the rate of change of 
	$G_{kj}$ with respect to $G_{kj'}$
	is increased by $\hat{\beta}_1$\%, 
	if the rate of change of $\hat{G}_{(k+1)j}$ with respect to $\hat{G}_{(k+1)j'}$ is increased by 1\%. According to the normal economic theory, the import structure should be stable and the overall change rate be 1 if there is no structural disturbance. If $|\hat{\beta}_1 -1| \gg 0$, the import structure will be greatly influenced.  Additionally, there is one issue deserving our attention: the change rate of $\hat{\beta}_1$ treats every component in the same speed, and to some degree reflects the correlation of different components. In other word, $\hat{\beta}_1$ is a comprehensive parameter without considering the heterogeneity among different components. Therefore, we will infuse the GMDGMSS with the TGMI model, which can fulfill this matter.

\subsection{TGMI Model}

TGMI established by Song in 2019 \cite{GMilr} is a model with {\textit{D-1}} sub-models, whose response and control coefficients are different from each other. The main process of TGMI lies in \emph{ilr} transformation before the modeling.
 
\begin{definition}
	$\forall \mathbf{X} \in S_n^D$, the ilr transformation ilr($\mathbf{X}$) is defined as,
	\begin{equation}
		ilr(\mathbf{X}) = 
		\begin{pmatrix}
			ilr(\mathbf{x}_1) \\
			ilr(\mathbf{x}_2) \\
			\vdots\\
			ilr(\mathbf{x}_n) \\
		\end{pmatrix}
		\triangleq \mathbf{U}
	\end{equation}
	where $\mathbf{U} = (u_{ij})_{n(D-1)}, i=1,2,...,n;j=1,2,...,D-1$.
	
\end{definition}

The transformation removes the constant-sum and non-negative constraints of $\mathbf{X}$, which makes it possible to build the GM(1,1) model with $\mathbf{U}$ as usual. Consequently,$D-1$ GM(1,1) models (denoted as $\Omega$) works on each $\mathbf{u}_i, i= 1,2,...,D-1$, where
\begin{gather}
\Omega_i : u_i^{(0)}(k) + a_iz_i^{(1)}(k) =  b_i,\quad k=2,...,n; i=1,2,...,D-1
\end{gather}
where $\Omega_i$ is the $i^{th}$ model, $z_i^{(1)}(k) = 0.5u_i^{(0)}(k) + 0.5 u_i^{(0)}(k-1)$, and $ \mathbf{U}^{(0)}$ is transformed from $\mathbf{X}$ by {\textit{ilr}}. The following modeling and its solution of TGMI is the same with that of traditional GM(1,1) \cite{GMilr}.

%
%
%
%
%
%

\subsection{IGADGM Modeling}
GADGMSS deals with the compositional time series as a unit and ensure the correlation of different dimensions. However, GADGMSS ignores the heterogeneity of these compositions. For example, in the 
Theorem.\ref{coef}, $	(\hat{\beta}_1)^T=(\mathbf{\dot{X}}^{(1)}_0,\mathbf{\dot{X}}^{(1)}_0)^{-1}_S(	\mathbf{\dot{X}}^{(1)}_0,	\mathbf{\dot{X}}^{(1)}_1)_S$ means $\hat{\beta}_1$ is a constant figure instead of a vector in description of the development among dimensions. 

TGMI is able to fill the gap of the GADGMSS, seeing that TGMI treats these dimensions with {\textit{D-1}} GM(1,1) models. Therefore, we propose IGADGM to inherit merits of both the GADGMSS and TGMI.    
\begin{equation}
	\hat{\mathbf{X}}_I=\frac{\delta_T}{\delta_T+\delta_S}\otimes\hat{\mathbf{X}}_S
	\oplus
	\frac{\delta_S}{\delta_T+\delta_S}\otimes\hat{\mathbf{X}}_T
\end{equation}
\noindent where $\delta_T,\delta_S$ is the Component Vector Percentage Error (CVPE), which is illustrated in Table.\ref{error_all}, of TGMI and GADGMSS respectively. $\hat{X}_T, \hat{X}_S$ is the prediction of the two models separately. $\delta_S/(\delta_T+\delta_S)$ is called Determination Factor (DF). For example, if the $DF = 0.5$, the part of prediction of GADGMSS determining the new model is equal to that of TGMI, that is, the homogeneity of crude oil import structure in different areas is equal to heterogeneity and the corresponding policy should adopt a balanced approach to crude oil imports from different regions. 

\subsection{General Matrix Optimized with DE Algorithm} 
The selection of $\mathbf{B}$ in GADGMSS can be transformed into the nonlinear optimization problem as follows, 
\begin{gather}
\mathop{min}\limits_{\mathbf{B}}(CVPE) = \mathop{min}\limits_{\mathbf{B}} \frac{\langle ilr(\hat{X}^{(0)}),ilr(X^{(0)}) \rangle}{\langle ilr(X^{(0)},ilr(X^{(0)}\rangle} \label{min} \\ 
s.t.
\begin{cases}
X^{(1)}=\mathbf{B}\otimes X^{(0)} \\
\hat{P}=(\mathbf{\dot{X}}^{(1)}_0,\mathbf{\dot{X}}^{(1)}_0)^{-1}_S(	\mathbf{\dot{X}}^{(1)}_0,	\mathbf{\dot{X}}^{(1)}_1)_S\\
\mathbf{\hat{\dot{X}}}^{(1)}_1 = \mathbf{\dot{X}}^{(1)}_0\otimes P\\
\mathbf{\hat{X}}^{(0)}=\mathbf{B}^{-1}\otimes\mathbf{\hat{X}}^{(1)}
\end{cases}
\end{gather}
where $\mathbf{B}$ is the General-Accumulation matrix with $\frac{(n+1)n}{2}$ coefficients. The selection of the matrix means extracting the valuable information from the raw data and it is crucial to the performance of GADGMSS and IGADGM.  Moreover, the number of the $\mathbf{B}$ increasing with the square of its samples, that is $O(n^2)$, makes it challenging to find the global optimal solution in a short time. Therefore, we adopt Best One-Dimension DE optimizer \cite{applied,overall}, which are represented in Algorithm.\ref{DE algorithm}. 

DE algorithm is one of the evolutionary algorithms(EAs), which are stochastic search method that mimic evolutionary process encountered in nature, and have been successfully applied to a wide range of optimization problems\cite{applied,Memodify}. The main four steps of the EAs are characterized as {\textit{Initialization, Mutation Operators, Crossover, and Selection}} \cite{overall}.

\begin{algorithm}[!htb]
	\SetAlgoLined
	\KwIn{Population: M; Dimension: D; Generation: T}  
	\KwOut{The best vector $\Delta$} 
	t  $\gets$ 1 
	
    $\triangleright$(initialization)
	
	\For{$i = 1 $ to $M$}  	
	{  
		\For{$j = 1$ to $D$}  
		{  
			$x^j_{i,t} = x^j_{min} + rand(0,1)*(x^j_{max}-x^j_{min})$;
		}  
	}  
	\While{$(|f(\Delta)|\ge \epsilon)$ or $(T\ge t)$}  
	{   
		
		$\triangleright$(Mutation and Crossover)
		
		\For{$i=1$ to $M$} 	
		{   
			
			\For{$j=1$ to $D$}  
			{  
				$v^j_{i,t} = Mutation(x^j_{i,t})$;\\
				$u^j_{i,t} = Crossover(x^j_{i,t},v^j_{i,t})$
			}  
			
		}
		
		$\triangleright$	(Greedy Selection)
		
		\eIf{$f(\mathbf{u}_{i,t})\leq f(\mathbf{u}_{i,t})$}{
			$\mathbf{x}_{i,t}\gets \mathbf{u}_{i,t}$;\\
			\If{$f(\mathbf{u}_{i,t})\leq f(\Delta)$}{	$\Delta \gets \mathbf{x}_{i,t}$}   	} {$\mathbf{x}_{i,t}\gets \mathbf{x}_{i,t}$}
		$t\gets t+1$
	}
	\Return{the best vector $\Delta$}
	\caption{Differential Evolution Algorithm (DE algorithm)} \label{DE algorithm}
\end{algorithm}

\section{Validation and Application}\label{sec3}

\subsection{Prepocessing Crude Oil Dataset}

The three energy import datasets are considered in this study: Canada's organic chemicals, India's crude oil and related energy, and China's crude oil and related energy, all of which derives from real statistical data\cite{dataset}. The first two datasets are utilized to validate the IGADGM's better performance over TGMI and GADGMSS. The last dataset is compared with currently popular models and used to conduct an empirical analysis as a scientific basis for future policy.

\begin{figure}[!htb]
	\centering
	\includegraphics[width=1\textwidth]{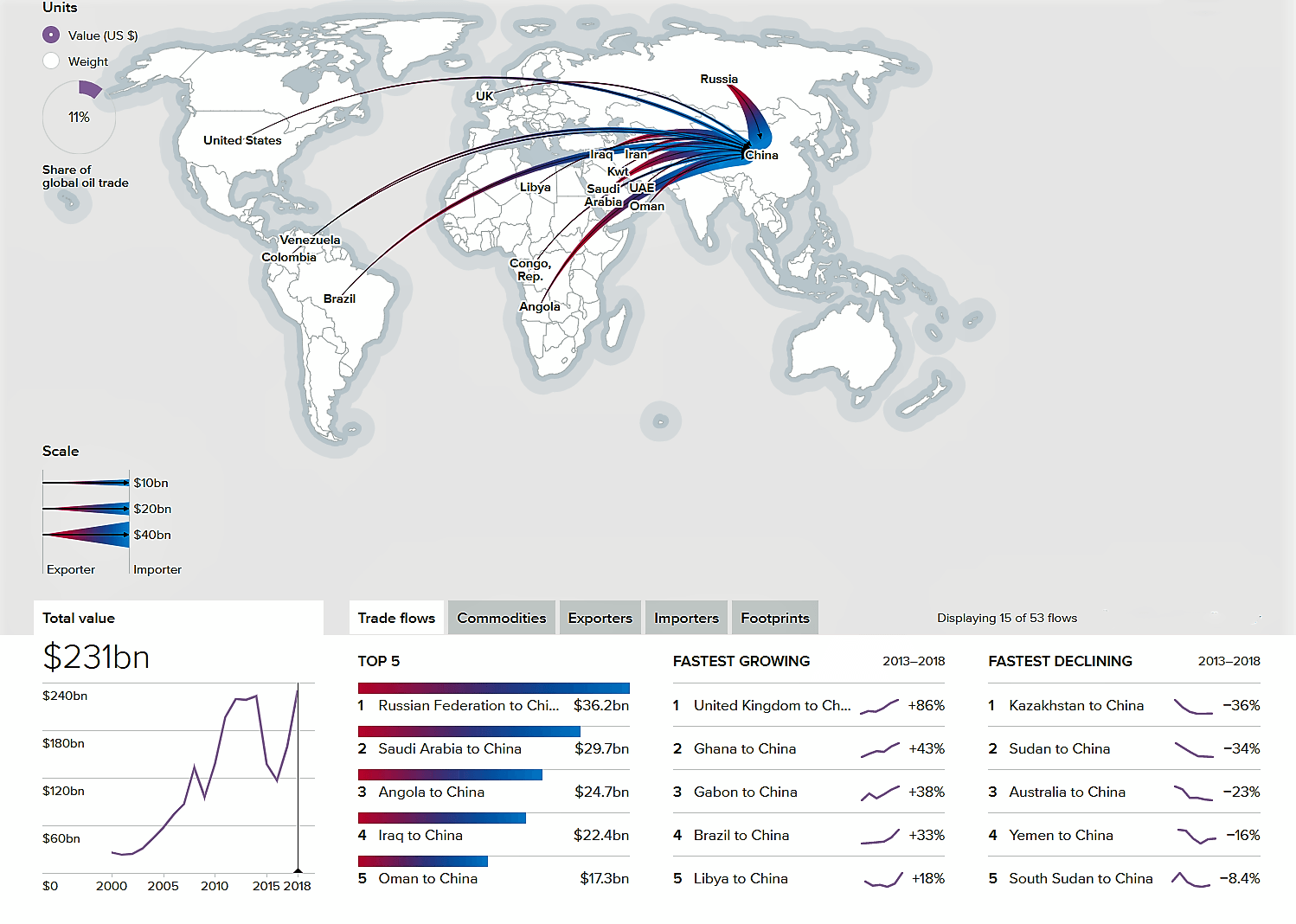} 
	\caption{The Distribution of the Crude Oil Import of China} 
	\label{imgo}
\end{figure}

Notably, an important data preprocessing needs to be stressed. The raw sequence is real value of import structure information, which cannot be directly regarded as the compositional data. Therefore, it is necessary for import sources be divided into different regions and the real value is then transformed into compositional data. To take the example of China's crude oil import dataset (other two dataset works as the same way), the real dataset is mapped out in the Figure.\ref{imgo}, where the data is characteristic of three types: limited sample (only 10 observations), many dimensions (over 70 countries), and centralized distribution. However, for the excessively high data dimensions, the first methods that come to mind are Principal Component Analysis (PCA) and Factor Analysis (FA), but these methods lose their power under the limited sample. Although there are approaches to solve this problem, for instance LASSO and t-SNE, the results from these quantitative methods cannot be well explained and therefore overestimate the significance of exporting countries in Arctic area thanks to the opening of Arctic shipping. Therefore, this paper adopts an empirical way to split the exporting countries into several regions, which satisfies appropriate distribution of different countries and  easy to explain the economic significance. 

The major sources of crude oil import in China is divided into five regions: Middle East, Africa, Southeast Asia, the Arctic regions, America\cite{patterns}. The processed data from 2010 to 2019 is shown in Table.\ref{china_data}.

\begin{table}[!htb]\footnotesize
	\centering
	\caption{Crude Oil Import Composition Data of China (2010 to 2019)}
	\label{china_data}
	\resizebox{\textwidth}{16mm}{
		\begin{tabular}{ccccccccccc}
			\toprule
			Region&2010&2011&2012&2013&2014&2015&2016&2017&2018&2019\\
			\midrule
			Arctic nations&10.23\%&12.06\%&13.63\%&12.81\%&13.38\%&15.15\%&16.67\%&19.12\%&19.77\%&18.29\%\\
			
			Middle east &46.77\%&49.50\%&49.75\%&51.76\%&52.79\%&50.38\%&47.01\%&43.56\%&45.75\%&47.34\%\\
			
			Africa&21.92\%&18.53\%&16.87\%&16.58\%&15.61\%&12.63\%&12.63\%&13.26\%&12.42\%&11.93\%\\
			
			America&8.28\%&8.46\%&9.18\%&8.42\%&9.42\%&10.10\%&10.81\%&10.92\%&11.03\%&11.28\%\\
			
			Southeast Asia&12.79\%&11.44\%&10.55\%&10.43\%&8.80\%&11.74\%&12.89\%&13.13\%&11.03\%&11.15\%\\
			\bottomrule
	\end{tabular}}
\end{table}

 Specifically, the five regions are: the arctic nations (Russian Federation, United States of America, Canada, Norway); the Middle East countries (Saudi Arabia, Iraq, Oman, Kuwait, United Arab Emirates, Qatar, Islamic Republic of Iran, Libya, Yemen); the African countries (Angola, Congo, Nigeria, Sudan, South Africa); the southeast Asian countries (Malaysia, Indonesia, Singapore, Thailand, Vietnam, Philippines); and the America.

\subsection{Metrics of Performance Evaluation}

The error metrics formulas in the following Table.\ref{error_all}.

\begin{table}[!htb]
\centering
\caption{Three Metrics on Models Effectiveness}
\label{error_all}
\setlength{\tabcolsep}{12mm}{
\begin{tabular}{ccc}
\toprule
Name                              & Abbreviation & Formula                                                                      \\ \hline
Component Vector Percentage Error & CVPE         & $\frac{\langle ilr(X_1),ilr(X_2) \rangle}{\langle ilr(X_1),ilr(X_1)\rangle}$ \\
Mean Component Percentage Error & MCPE & $\frac{1}{n}\sum_{i=1}^{n}\frac{\langle ilr(x_{1i}),ilr(x_{2i}) \rangle}{\langle ilr(x_{1i}),ilr(x_{1i})\rangle}$ \\
Mean Absolute Percentage Error    & MAPE         & $\frac{1}{nD}\sum_{i=1}^{n}|\frac{ilr(x_{1i})-ilr(x_{2i})}{ilr(x_{1i})}|$    \\ \bottomrule
\end{tabular}}
\end{table}

\subsection{Validation of Grey models}

In this section, the effectiveness and robustness of IGADGM is validated, that being said, we would testify whether the improvement on TGMI and GADGMSS is working well or not.  All grey prediction models, including the new proposed model and the two grey contrast models, are implemented in Python. The following two examples use the two different splits of the datasets. Specifically, 10 years are separated to 7 years data for training and the rest 3 years for testing (abbreviated as 7/3). That way as well applies to the 8/2.
 
\subsubsection{Example A: The Imports of Organic Chemicals in Canada (8/2)}
Import sources are divided into three regions:  America, Europe, and Asia and shown in Table.\ref{c_data}. 
\begin{table}[!htb]
	\footnotesize 
	\centering 
	\caption{Organic Chemicals Import Composition Data of Canada (2010 to 2019)}
	\label{c_data}
	\resizebox{\textwidth}{!}{
		\begin{tabular}{ccccccccccc}
			\toprule
			Region&2010&2011&2012&2013&2014&2015&2016&2017&2018&2019\\
			\midrule
			America&57.32\%&59.67\%&59.55\%&60.33\%&63.56\%&56.55\%&58.27\%&57.03\%&47.81\%&44.08\%\\
			
			Europe&26.69\%&24.82\%&23.43\%&23.78\%&20.91\%&27.66\%&24.97\%&24.72\%&34.45\%&36.33\%\\
			
			Asia&15.99\%&15.51\%&17.03\%&15.89\%&15.53\%&15.79\%&16.75\%&18.26\%&17.74\%&19.59\%\\
			\bottomrule
	\end{tabular}}
\end{table}

\begin{figure}[!htb]
	\centering
	\includegraphics[width=.7\textwidth]{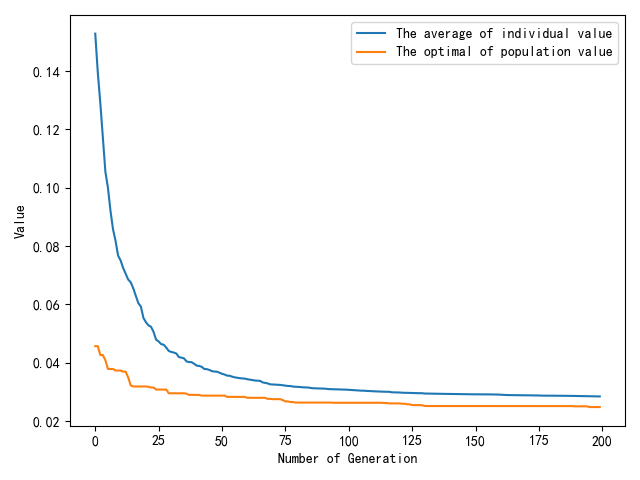} 
	\caption{The Convergence of DE algorithm with 200 generations}
	\label{img15}
\end{figure}

The parameter is calculated $\hat{\beta}_1 = 0.537$. The process of DE algorithm evolves in Figure.\ref{img15}. The short vertical yellow lines (or steep slope lines) mean skipping out of the local optimal solution. Clearly, the yellow line finally tends to be horizontal before several of the special lines, indicating that the global optimal solution is likely to be obtained. That is to say, the DE algorithm works well under such situation.

\begin{figure}[!htb]
	\centering
	\includegraphics[width=.95\textwidth]{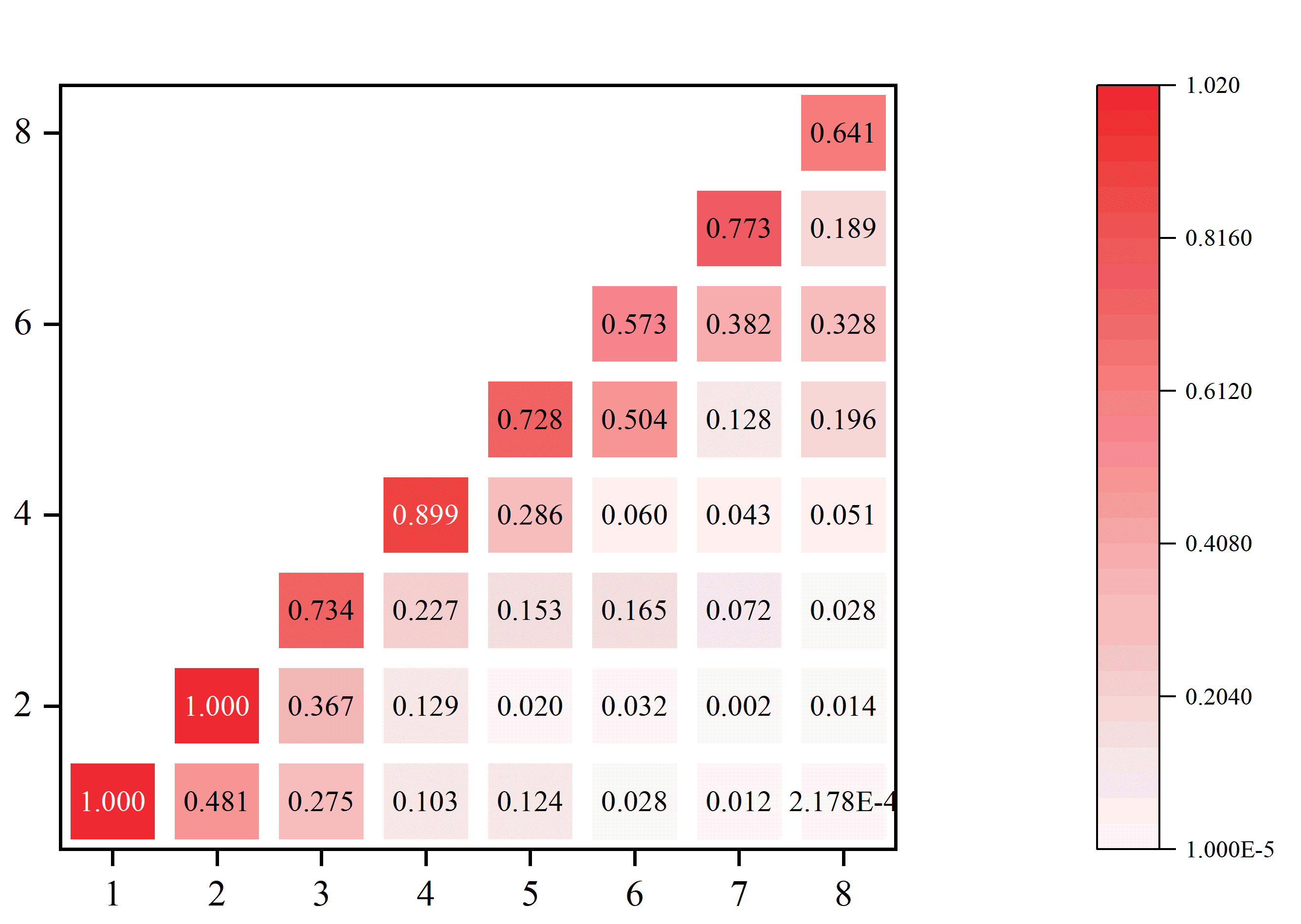} 
	\caption{General-Accumulation Matrix: Coefficient with Deeper Color Weighs More Importantly} 
	\label{img1}
\end{figure}

 Moreover, the matrix $\mathbf{B}$ optimized by the DE algorithm is shown as follows in Figure.\ref{img1}. The darker the red is, the larger current compositions' weight is, which turns out to satisfy the principle of information priority. 

\begin{figure}[!htb]
	\centering
	\includegraphics[width=.75\textwidth]{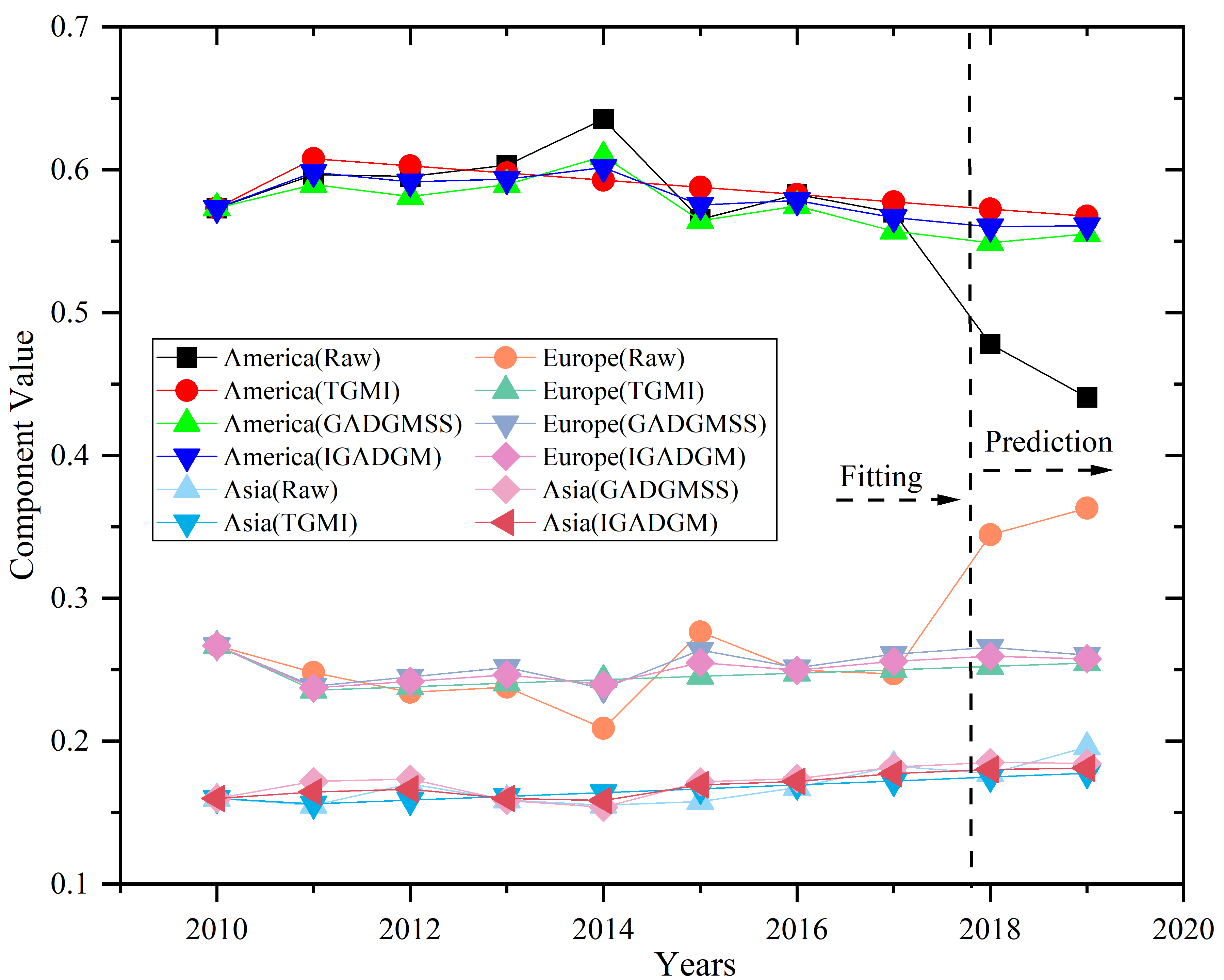} 
	\caption{Three Models on Canada's Chemical Import Structure} 
	\label{Can_pred}
\end{figure}

The three models' fitting and prediction is shown in Table.\ref{Canda_predata} and Figure.\ref{Can_pred}. Basically, the three models seem no significant difference, if one model has a better accuracy of prediction at one region, the model will has a worse prediction at another one. It is, however, demanding to observe which one works best, and thus this paper adopts three types of error evaluation to select the best, which are shown in Table.\ref{error_all}
\begin{table}[!htb]
	\centering
	\caption{The Prediction on Canada's Improt Percentage ($\times 100\%$) }
	\label{Canda_predata}
	\resizebox{\textwidth}{!}
	{
		\begin{tabular}{cccccccccc}
			\toprule
			& \multicolumn{3}{c|}{American}                & \multicolumn{3}{c|}{Europe}                  & \multicolumn{3}{c}{Asia} \\ \cline{2-10} 
			Times & TGMI & GADGMSS & \multicolumn{1}{c|}{IGADGM} & TGMI & GADGMSS & \multicolumn{1}{c|}{IGADGM} & TGMI  & GADGMSS & IGADGM \\ \hline
			2010 & 0.57 & 0.57 & 0.57 & 0.27 & 0.27 & 0.27 & 0.16 & 0.16 & 0.16 \\
			2011 & 0.61 & 0.59 & 0.60 & 0.24 & 0.24 & 0.24 & 0.16 & 0.17 & 0.16 \\
			2012 & 0.60 & 0.58 & 0.59 & 0.24 & 0.25 & 0.24 & 0.16 & 0.17 & 0.17 \\
			2013 & 0.60 & 0.59 & 0.59 & 0.24 & 0.25 & 0.25 & 0.16 & 0.16 & 0.16 \\
			2014 & 0.59 & 0.61 & 0.60 & 0.24 & 0.24 & 0.24 & 0.16 & 0.15 & 0.16 \\
			2015 & 0.59 & 0.56 & 0.58 & 0.25 & 0.26 & 0.26 & 0.17 & 0.17 & 0.17 \\
			2016 & 0.58 & 0.57 & 0.58 & 0.25 & 0.25 & 0.25 & 0.17 & 0.17 & 0.17 \\
			2017 & 0.58 & 0.56 & 0.57 & 0.25 & 0.26 & 0.26 & 0.17 & 0.18 & 0.18 \\
			2018 & 0.57 & 0.55 & 0.56 & 0.25 & 0.27 & 0.26 & 0.17 & 0.19 & 0.18 \\
			2019 & 0.57 & 0.56 & 0.56 & 0.25 & 0.26 & 0.26 & 0.18 & 0.18 & 0.18 \\ \bottomrule
		\end{tabular}
	}
\end{table}

\begin{figure}[H]
	\centering
	\includegraphics[width=.85\textwidth]{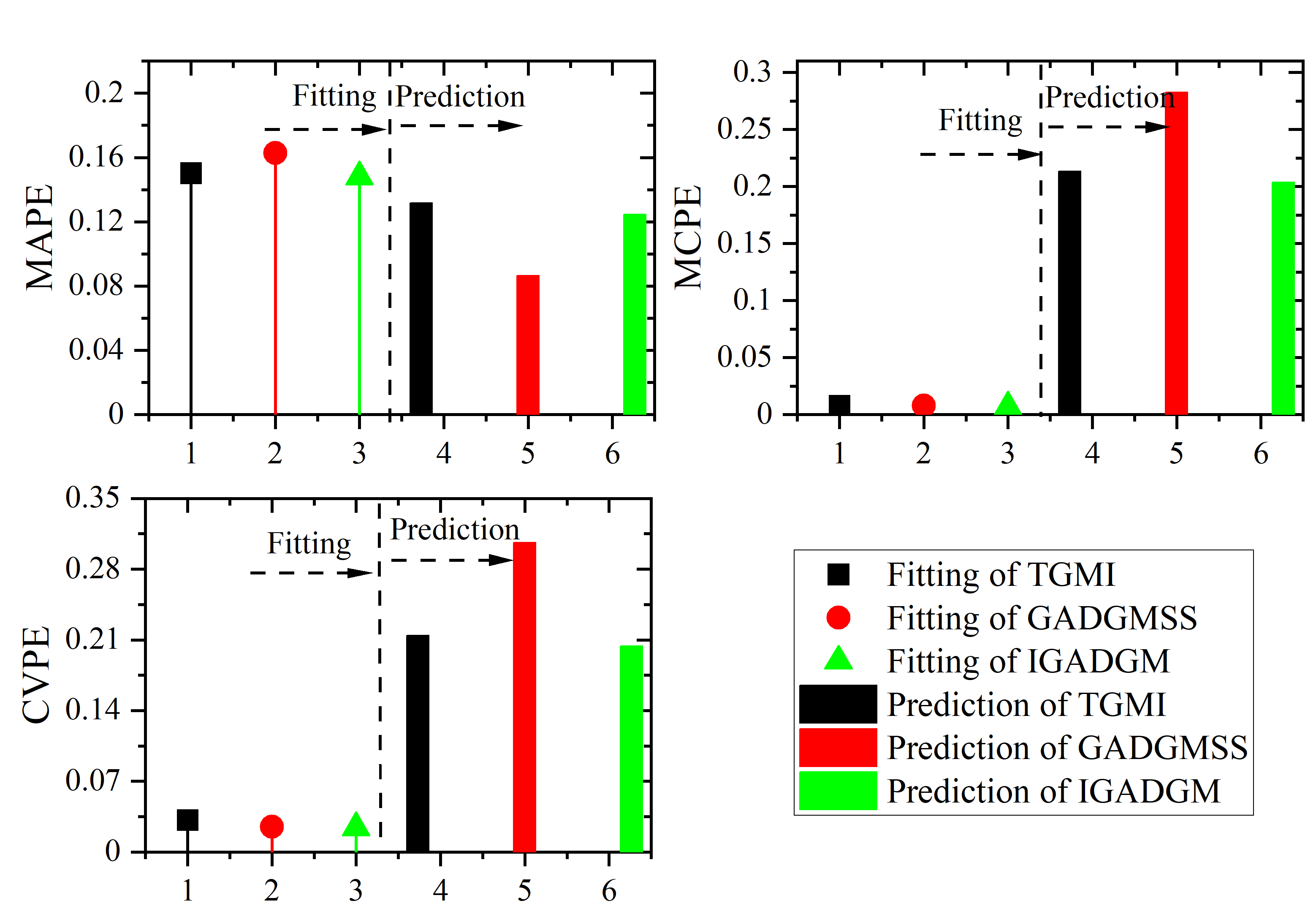} 
	\caption{The Error Evaluation of Three Models on Canada's Chemical Import Structure} 
	\label{CanError_pred}
\end{figure}

The DF of IGADGM is 0.5882, which means the contribution of GADGMSS accounts for 58.82\% of eventual prediction. That being said, the homogeneity Canada import structure on organic chemicals is slightly exceeding than its heterogeneity. Moreover, the minimum errors are bold in the Table.\ref{error_canda}, from which IGADGM has an advantage over others on the CVPE and MCPE, except on the MAPE. From the fitting error, the IGADGM exceeds the other two models among the three evaluations.

\begin{table}[H]
	\centering
	\caption{Three Types of Error Evaluating Three Models Performance from Fitting and Prediction Respectively (8/2) ($\times 100\%$)}
	\label{error_canda}
	\setlength{\tabcolsep}{7mm}{
		\begin{tabular}{ccccccc}
			\toprule
			& \multicolumn{3}{c|}{Prediction Error}   & \multicolumn{3}{c}{Fitting Error}                \\ \cline{2-7} 
			& CVPE  & MCPE  & MAPE           & CVPE  & MCPE  & MAPE  \\ \hline
			TGMI    & 0.215 & 0.213 & 0.132          & 0.032 & 0.008 & 0.150 \\
			GADGMSS & 0.307 & 0.283 & \textbf{0.087} & 0.025 & 0.007 & 0.163 \\
			IGADGM & \textbf{0.204} & \textbf{0.204} & 0.125 & \textbf{0.024} & \textbf{0.006} & \textbf{0.148} \\ \bottomrule
		\end{tabular}
	}
\end{table}

\subsubsection{Example B: The Imports of Crude Oil in India (7/3)}

\begin{table}[H]\footnotesize
	\centering
	\caption{Crude Oil Import Composition Data of India (2010 to 2019)}
	\label{India_data}
	\resizebox{\textwidth}{!}{
		\begin{tabular}{ccccccccccc}
			\toprule
			Region&2010&2011&2012&2013&2014&2015&2016&2017&2018&2019\\
			\midrule
			Middle East&58.20\%&61.28\%&60.64\%&59.96\%&57.30\%&54.13\%&55.91\%&55.05\%&57.04\%&54.29\%\\
			
			Africa&19.26\%&16.79\%&15.52\%&14.50\%&15.64\%&17.67\%&15.52\%&14.36\%&13.56\%&14.85\%\\
			
			America&8.57\%&8.03\%&12.64\%&15.02\%&14.71\%&12.09\%&11.00\%&12.07\%&13.14\%&14.23\%\\

			Asia-Pacific&13.97\%&13.90\%&11.20\%&10.52\%&12.35\%&16.12\%&17.57\%&18.52\%&16.27\%&16.63\%\\
			
			\bottomrule
	\end{tabular}}
\end{table}

\begin{table}[H]
	\centering
	\caption{The Prediction on India's Improt Percentage ($\times 100\%$)}
	\label{India_predata}
	\resizebox{\textwidth}{!}{
	\begin{tabular}{ccccccccccccc}
		\toprule
		&
		\multicolumn{3}{c|}{Middle   East} &
		\multicolumn{3}{c|}{Africa} &
		\multicolumn{3}{c|}{America} &
		\multicolumn{3}{c}{Asia\&Pacific} \\ \cline{1-13} 
		Times &
		TGMI &
		GADGMSS &
		\multicolumn{1}{c|}{IGADGM} &
		TGMI &
		GADGMSS &
		\multicolumn{1}{c|}{IGADGM} &
		TGMI &
		GADGMSS &
		\multicolumn{1}{c|}{IGADGM} &
		TGMI &
		GADGMSS &
		IGADGM \\ \cline{2-13} 
		2010 & 0.58 & 0.58 & 0.58 & 0.19 & 0.19 & 0.19 & 0.09 & 0.09 & 0.09 & 0.14 & 0.14 & 0.14 \\
		2011 & 0.62 & 0.62 & 0.62 & 0.16 & 0.16 & 0.16 & 0.11 & 0.09 & 0.10 & 0.11 & 0.12 & 0.12 \\
		2012 & 0.61 & 0.57 & 0.58 & 0.16 & 0.18 & 0.17 & 0.11 & 0.11 & 0.11 & 0.12 & 0.15 & 0.14 \\
		2013 & 0.59 & 0.58 & 0.59 & 0.16 & 0.16 & 0.16 & 0.12 & 0.15 & 0.14 & 0.13 & 0.12 & 0.12 \\
		2014 & 0.58 & 0.56 & 0.57 & 0.16 & 0.16 & 0.16 & 0.12 & 0.16 & 0.14 & 0.14 & 0.12 & 0.13 \\
		2015 & 0.56 & 0.54 & 0.55 & 0.16 & 0.18 & 0.17 & 0.13 & 0.13 & 0.13 & 0.15 & 0.15 & 0.15 \\
		2016 & 0.55 & 0.55 & 0.55 & 0.16 & 0.18 & 0.17 & 0.13 & 0.12 & 0.12 & 0.16 & 0.16 & 0.16 \\
		2017 & 0.53 & 0.56 & 0.55 & 0.16 & 0.17 & 0.16 & 0.14 & 0.12 & 0.13 & 0.17 & 0.16 & 0.16 \\
		2018 & 0.52 & 0.55 & 0.54 & 0.16 & 0.18 & 0.17 & 0.14 & 0.13 & 0.13 & 0.18 & 0.15 & 0.16 \\
		2019 & 0.50 & 0.55 & 0.53 & 0.16 & 0.18 & 0.17 & 0.15 & 0.13 & 0.14 & 0.19 & 0.14 & 0.16 \\ \bottomrule
	\end{tabular}
}
\end{table}

\begin{table}[H]
	\centering
	\caption{Three Types of Error Evaluating Three Models Performance from Fitting and Prediction Respectively (7/3) ($\times 100\%$)}
	\label{error_india}
	\setlength{\tabcolsep}{7mm}{
		\begin{tabular}{ccccccc}
			\toprule
			& \multicolumn{3}{c|}{Prediction Error}            & \multicolumn{3}{c}{Fitting Error}                \\ \cline{2-7} 
			& \multicolumn{1}{c}{CVPE} & \multicolumn{1}{c}{MCPE} & \multicolumn{1}{c}{MAPE} & \multicolumn{1}{c}{CVPE} & \multicolumn{1}{c}{MCPE} & \multicolumn{1}{c}{MAPE} \\ \hline
			TGMI    & 0.032          & 0.032          & \textbf{0.018} & 0.075          & 0.032          & 0.245          \\
			GADGMSS & 0.053          & 0.053          & 0.025          & \textbf{0.038} & \textbf{0.017} & \textbf{0.168} \\
			IGADGM  & \textbf{0.031} & \textbf{0.031} & 0.019          & 0.041          & 0.018          & 0.182  \\ \bottomrule
		\end{tabular}
	}
\end{table}

The preprocessed data shown in Table.\ref{India_data}. The parameter is calculated $\hat{\beta}_1 = 0.476$. The $DF = 0.6135$. The analysis of each figure and table is omitted.

\begin{figure}[H]
	\centering
	\includegraphics[width=.7\textwidth]{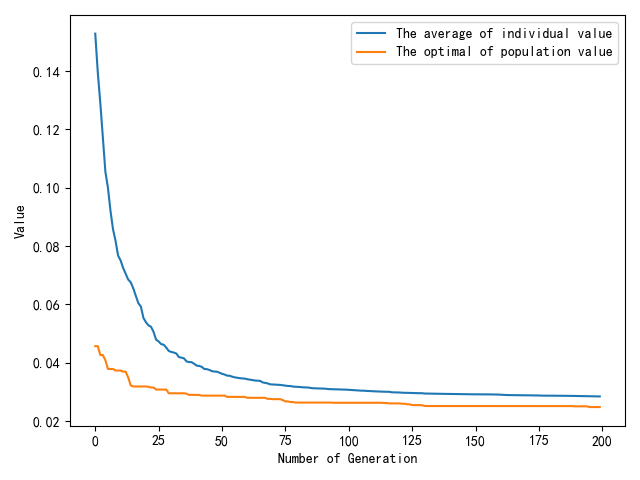} 
	\caption{The Convergence of DE algorithm with 200 generations}
\end{figure}


\begin{figure}[H]
	\centering
	\includegraphics[width=.7\textwidth]{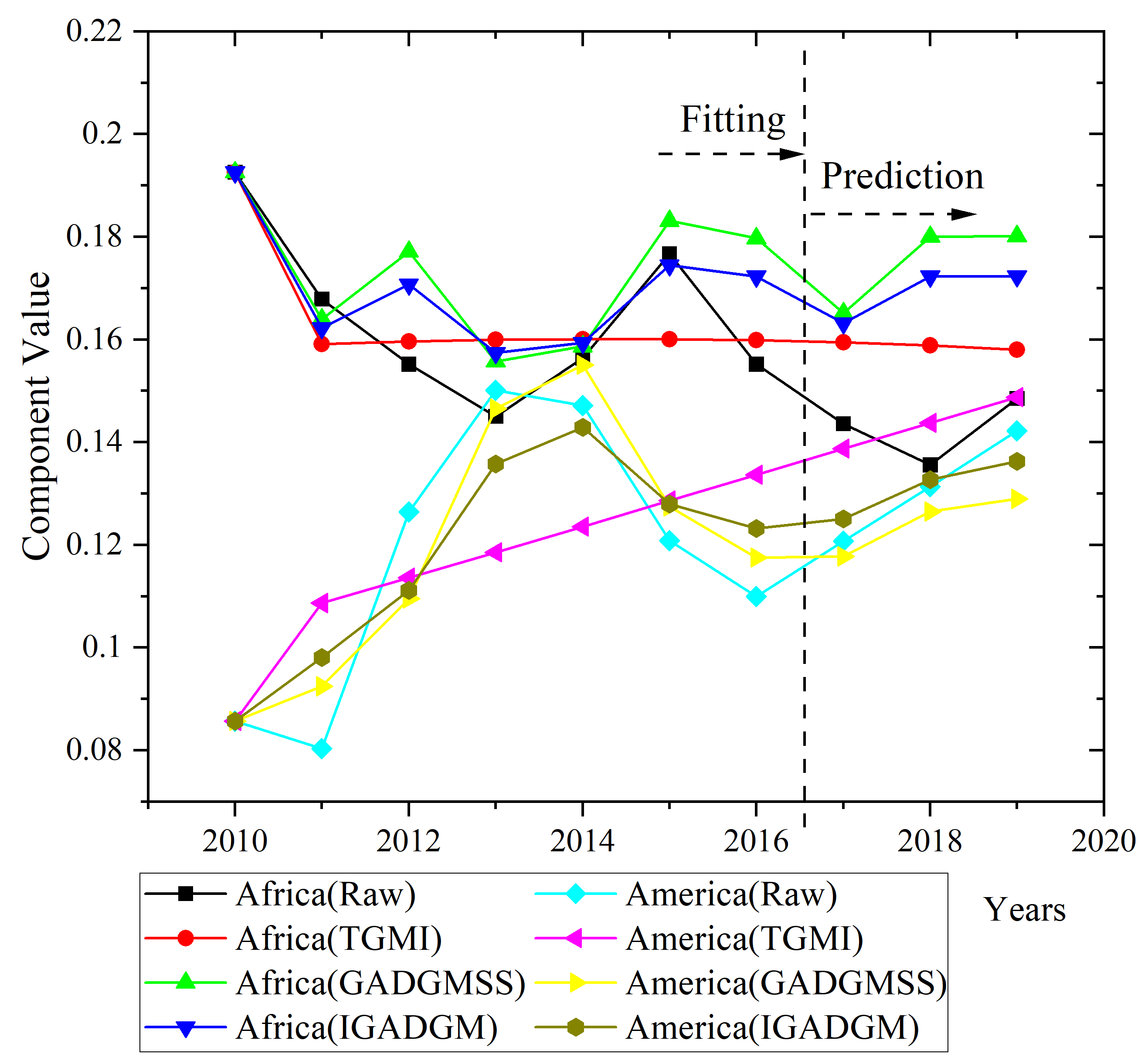} 
	\caption{Three Models on India's Crude Oil Import Structure (Africa and America)} 
\end{figure}

\begin{figure}[H]
	\centering
	\includegraphics[width=.7\textwidth]{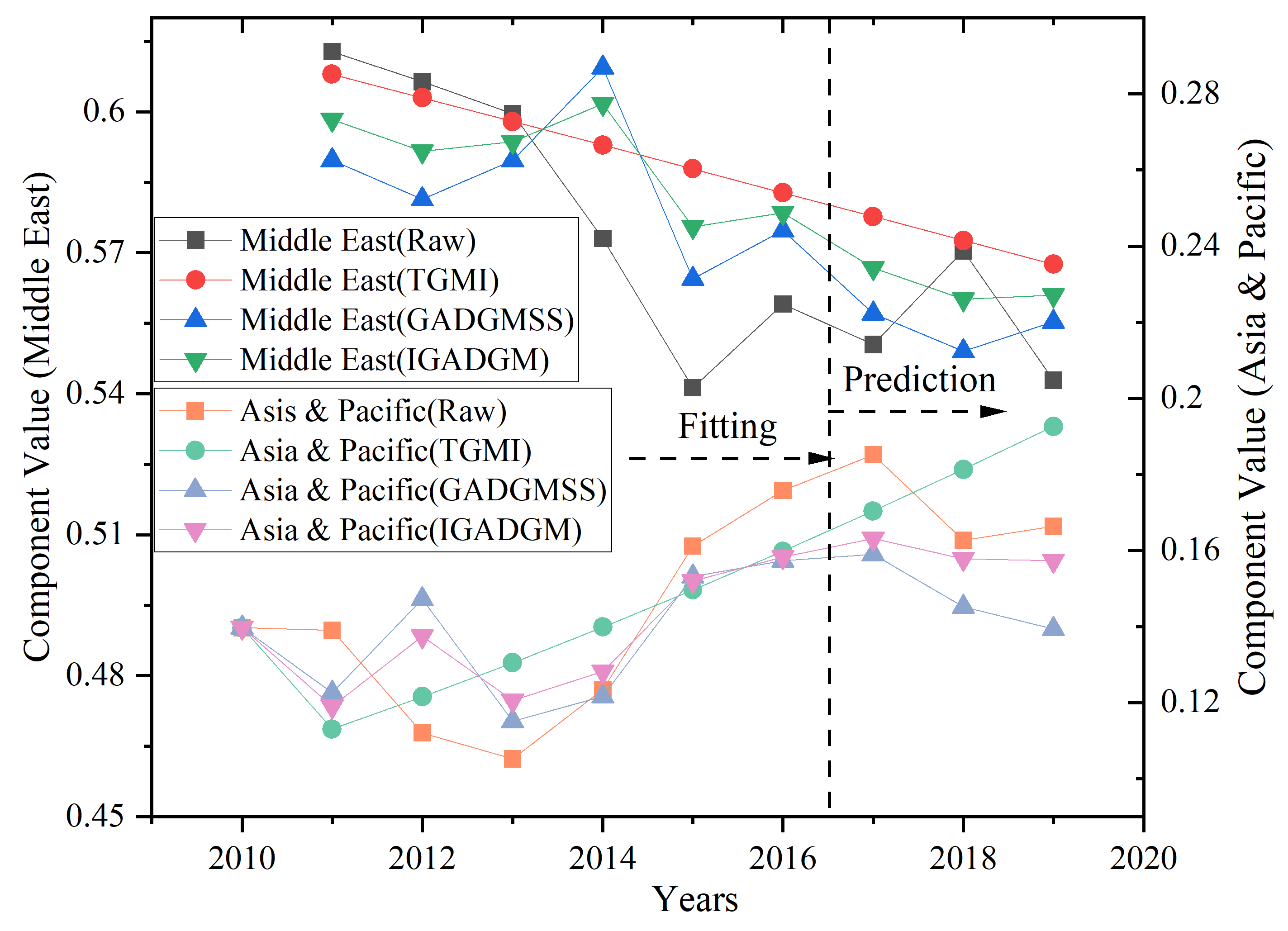} 
	\caption{Three Models on India's Oil Import Structure (Middle East and Asia \& Pacific)} 
\end{figure}


\begin{figure}[H]
	\centering
	\includegraphics[width=.85\textwidth]{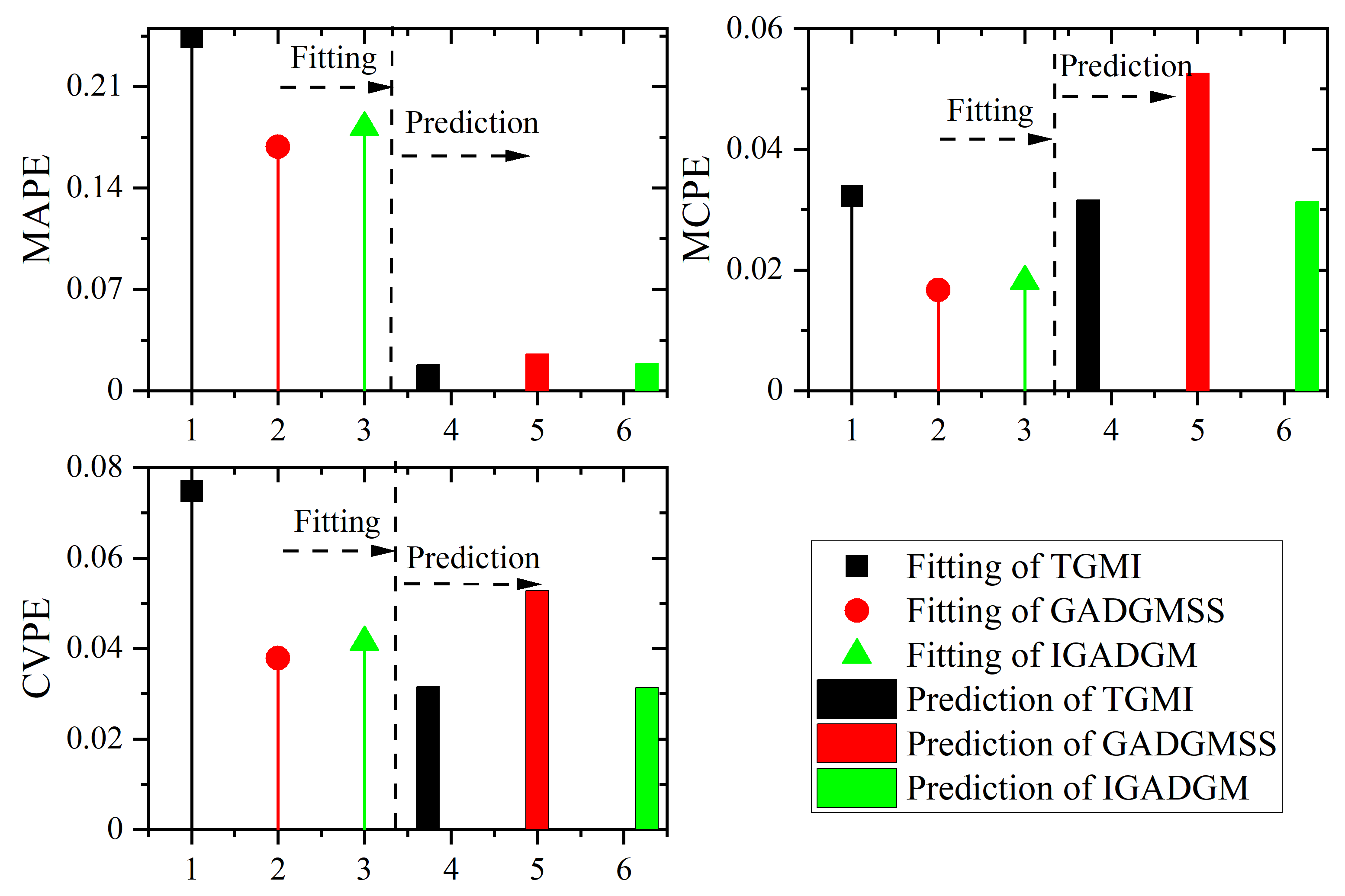} 
	\caption{The Error Evaluation of Three Models on India's Crude Oil Import Structure} 
\end{figure}

\subsubsection{Analysis and Discussion}

From the results of the two validation experiments, it can be noticed that the proposed IGADGM model has advantages over the other two models. 

Firstly, IGADGM has better robustness than models TGMI and GADGMSS. It can be seen from the above two examples that in Example A, the error of IGADGM is almost the smallest in all three evaluation indexes. In Example B, although the IGADGM model error is not the smallest, it is only less than 1\% larger than the smallest error. Such results may be due to the IGADGM model being able to combine the results through fitting errors, so that the model does not appear a significant deviation. Therefore, IGADGM can better adapt to different scenarios.

Secondly, IGADGM has better performance on abnormal datasets. In Example A, the data set is obviously abnormal. In the past two years, the exporting regions have undergone tremendous changes, so GADGMSS prioritizing regional homogeneity cannot be well predicted. However, IGADGM, which weighs homogeneity and heterogeneity according to models' error, achieved the smallest error with regard to CVPE and MCPE. This result may be due to the fact that DF is significantly less than 1, which means that IGADGM inherits most of the content of TGMI, thus ensuring the differences in many regions. Therefore, the abnormal data in the past two years has been fully considered.

Above all, it can be drawn that the improvement on TGMI and GADGMSS are effective. Meanwhile, IGADGM can well adapt to the scenarios if there is no knowledge which one model is better than another. By the way, two kinds of merchandise import structure prediction are involved in Example A and B, to some extent, signifying that the IGADGM model has the potential of merchandise import structure prediction.

\subsection{Practical Application of Novel Model and Analysis}
\subsubsection{Practical Application}

\begin{table}[H]
	\centering
	\caption{The Prediction of Five Models on China Crude Oil Import (*100\%)}
	\tiny
	\label{chinaTable}
	\resizebox{\textwidth}{!}
	{
	\begin{tabular}{ccccccccccc}
		\toprule
		Models     & 2010 & 2011 & 2012 & 2013 & 2014 & 2015 & 2016 & 2017 & 2018 & 2019 \\ \hline
		\multicolumn{11}{c}{Arctic}                                                      \\
		IGADGM     & 10.2 & 12.2 & 13.1 & 13.3 & 13.7 & 14.6 & 15.1 & 15.4 & 15.8 & 16.3 \\
		ARIMA      & 12.8 & 11.8 & 12.5 & 12.9 & 12.9 & 13.1 & 13.6 & 13.5 & 13.4 & 13.3 \\
		LR         & 10.9 & 11.7 & 12.4 & 13.3 & 14.1 & 14.9 & 15.8 & 16.7 & 17.6 & 18.5 \\
		SVM        & 11.7 & 12.4 & 12.8 & 12.9 & 13.3 & 13.8 & 13.1 & 12.8 & 12.8 & 12.8 \\
		GaussianNB & 10.3 & 12.1 & 13.7 & 12.9 & 13.4 & 15.2 & 14.5 & 14.5 & 14.5 & 14.5 \\
		\multicolumn{11}{c}{Middle East}                                                 \\
		IGADGM     & 46.8 & 49.3 & 49.5 & 50.6 & 51.3 & 50.9 & 51.0 & 50.6 & 50.7 & 50.9 \\
		ARIMA      & 50.2 & 46.4 & 49.2 & 50.5 & 50.8 & 51.3 & 53.1 & 52.8 & 52.3 & 51.9 \\
		LR         & 47.9 & 49.0 & 49.9 & 50.8 & 51.5 & 52.2 & 52.7 & 53.1 & 53.5 & 53.7 \\
		SVM        & 46.3 & 48.6 & 49.8 & 50.1 & 51.5 & 52.8 & 50.9 & 49.9 & 49.9 & 49.9 \\
		GaussianNB & 46.7 & 49.1 & 49.5 & 51.5 & 52.6 & 50.0 & 47.7 & 47.7 & 47.7 & 47.7 \\
		\multicolumn{11}{c}{Africa}                                                      \\
		IGADGM     & 21.9 & 18.8 & 17.4 & 16.3 & 15.2 & 14.2 & 13.1 & 12.9 & 12.2 & 11.4 \\
		ARIMA      & 17.0 & 21.6 & 18.4 & 16.6 & 16.6 & 15.6 & 13.3 & 14.0 & 14.6 & 15.1 \\
		LR         & 21.2 & 19.4 & 17.7 & 16.1 & 14.6 & 13.2 & 12.0 & 10.8 & 9.7  & 8.7  \\
		SVM        & 20.6 & 18.3 & 17.2 & 17.1 & 16.0 & 13.9 & 15.8 & 17.0 & 17.1 & 17.1 \\
		GaussianNB & 21.9 & 18.6 & 16.9 & 16.8 & 15.7 & 12.8 & 17.2 & 17.2 & 17.2 & 17.2 \\
		\multicolumn{11}{c}{America}                                                     \\
		IGADGM     & 8.3  & 8.6  & 9.0  & 9.1  & 9.3  & 10.0 & 10.2 & 10.5 & 10.8 & 11.0 \\
		ARIMA      & 9.1  & 9.0  & 8.9  & 9.0  & 8.8  & 9.1  & 9.3  & 9.1  & 9.0  & 9.0  \\
		LR         & 8.2  & 8.5  & 8.8  & 9.1  & 9.5  & 9.7  & 10.0 & 10.3 & 10.5 & 10.8 \\
		SVM        & 9.5  & 9.4  & 9.4  & 9.4  & 9.4  & 9.2  & 9.3  & 9.4  & 9.4  & 9.4  \\
		GaussianNB & 8.3  & 8.5  & 9.3  & 8.5  & 9.5  & 10.2 & 8.7  & 8.7  & 8.7  & 8.7  \\
		\multicolumn{11}{c}{South Asia}                                                  \\
		IGADGM     & 12.8 & 11.1 & 11.0 & 10.7 & 10.5 & 10.3 & 10.6 & 10.6 & 10.5 & 10.4 \\
		ARIMA      & 11.0 & 11.2 & 11.0 & 10.9 & 10.9 & 10.9 & 10.7 & 10.7 & 10.7 & 10.8 \\
		LR         & 11.9 & 11.5 & 11.1 & 10.7 & 10.3 & 9.9  & 9.5  & 9.1  & 8.7  & 8.3  \\
		SVM        & 12.0 & 11.3 & 10.9 & 10.5 & 9.8  & 10.4 & 10.8 & 10.9 & 10.9 & 10.9 \\
		GaussianNB & 12.8 & 11.6 & 10.7 & 10.5 & 8.9  & 11.9 & 12.0 & 12.0 & 12.0 & 12.0 \\ \bottomrule
	\end{tabular}}
\end{table}

In this section, we primarily compare IGADGM with other popular prediction models to have an experiment whether the proposed novel model has a better performance over others. Furthermore, the selected model will be implemented for predicting the import structure of crude oil in China. All prediction models, including the new proposed models ($DF = 0.5998$ and development parameter is 0.521) and the four contrast models, are implemented in Python. The four contrast models are ARIMA, Linear Regression (LR), SVM, and GaussianNB, which are all based on the \emph{ilr} transformation and implemented in Python. 

\begin{figure}[H]
	\centering
	\includegraphics[width=.65\textwidth]{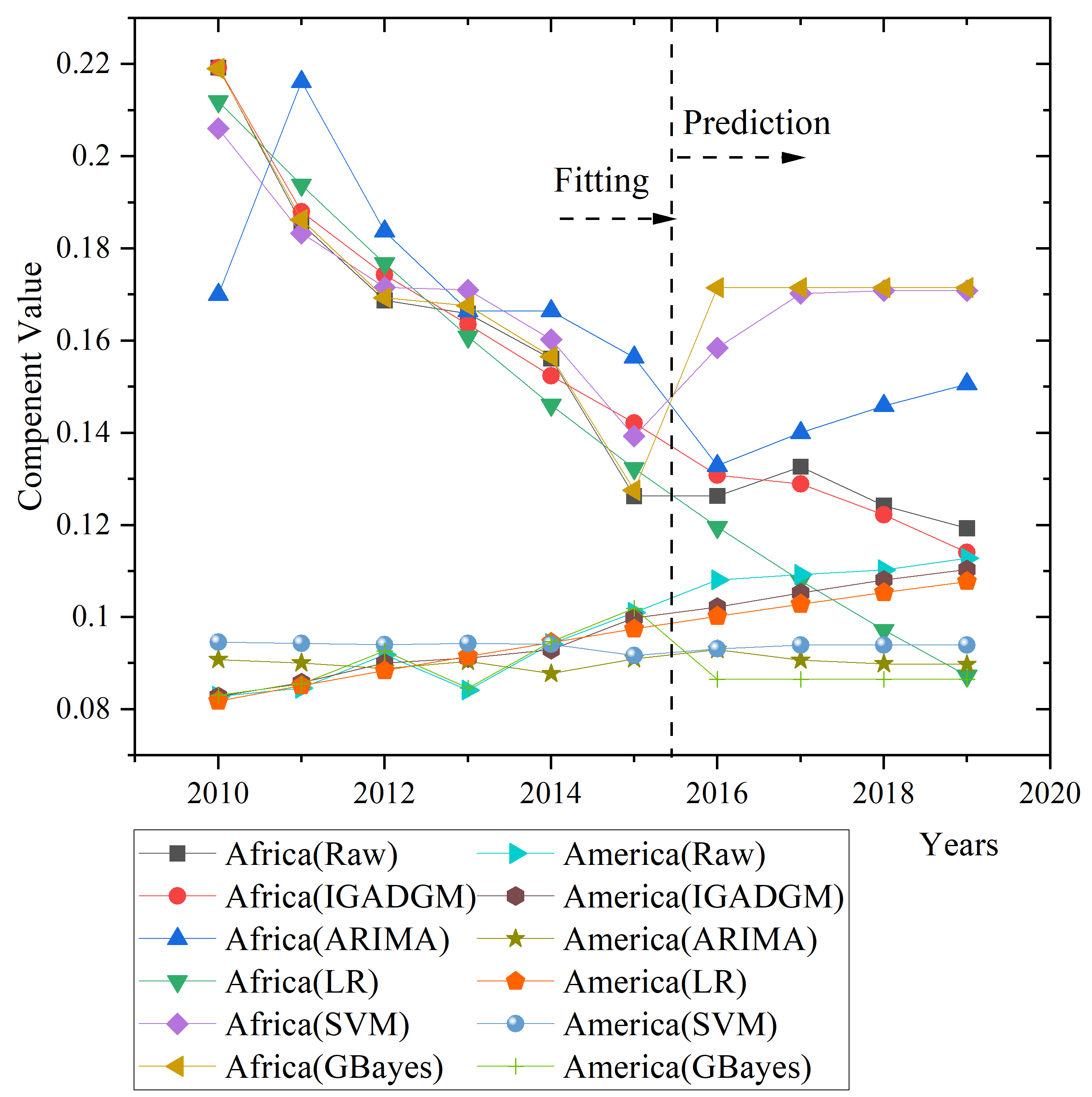} 
	\caption{Three Models on Arctic's Crude Oil Import Structure (Africa and America)} 
	\label{ar1}
\end{figure}
\begin{figure}[H]
	\centering
	\includegraphics[width=.65\textwidth]{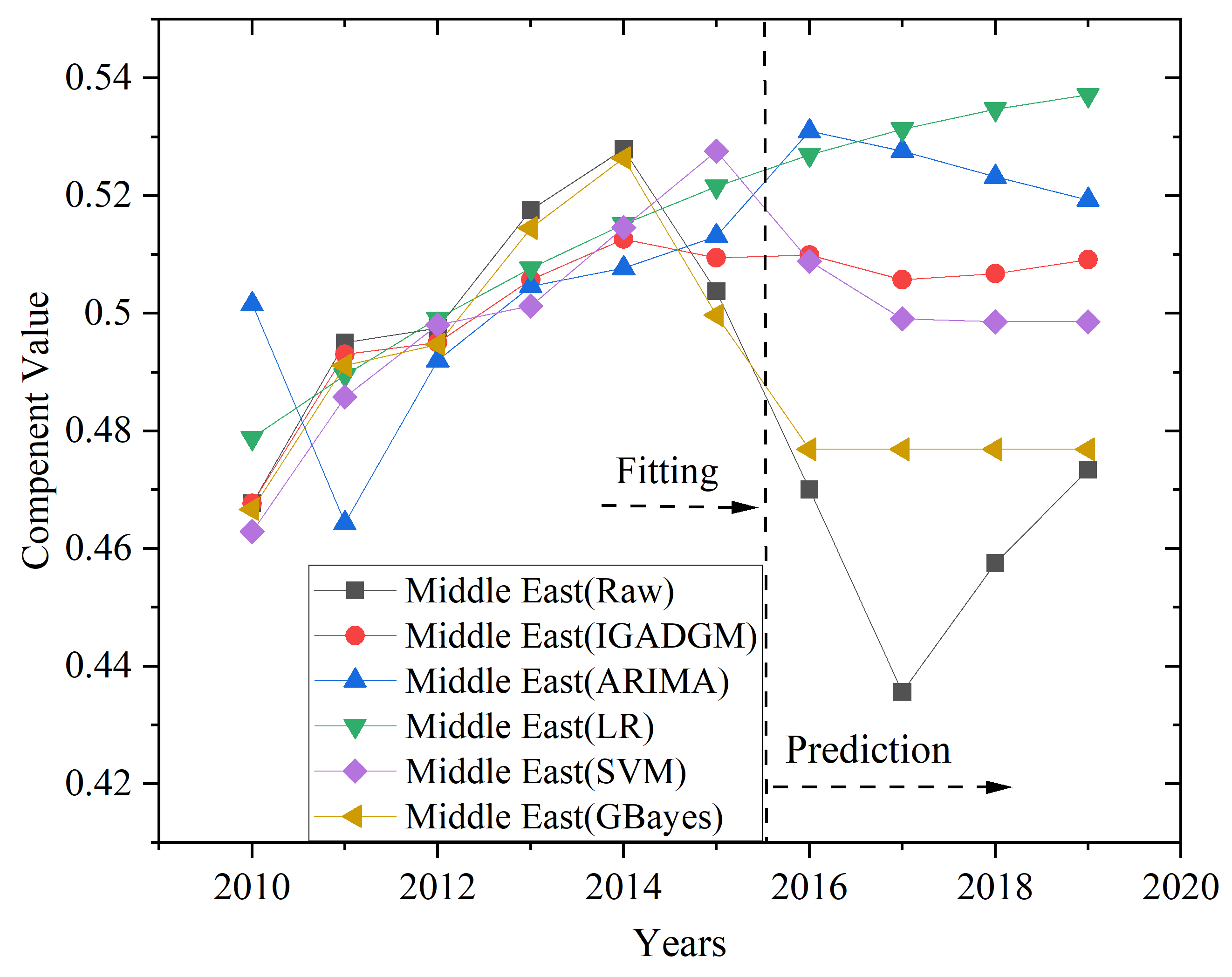} 
	\caption{Three Models on Arctic's Crude Oil Import Structure (Middle East)} 
	\label{ar3}
\end{figure}
\begin{figure}[H]
	\centering
	\includegraphics[width=.65\textwidth]{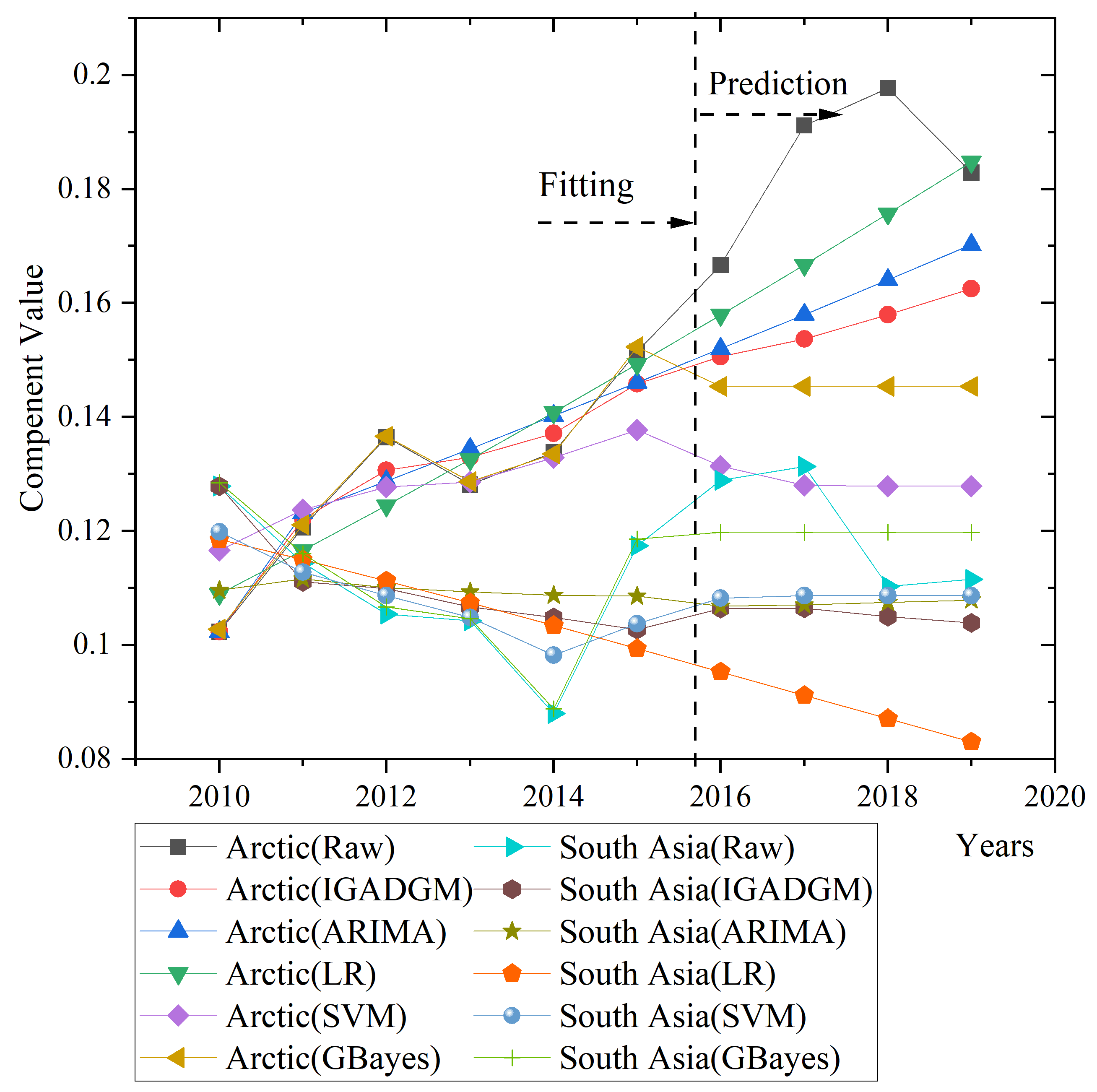} 
	\caption{Three Models on Arctic's Crude Oil Import Structure (Arctic and South Asia)} 
	\label{ar2}
\end{figure}
\begin{figure}[H]
	\centering
	\includegraphics[width=.85\textwidth]{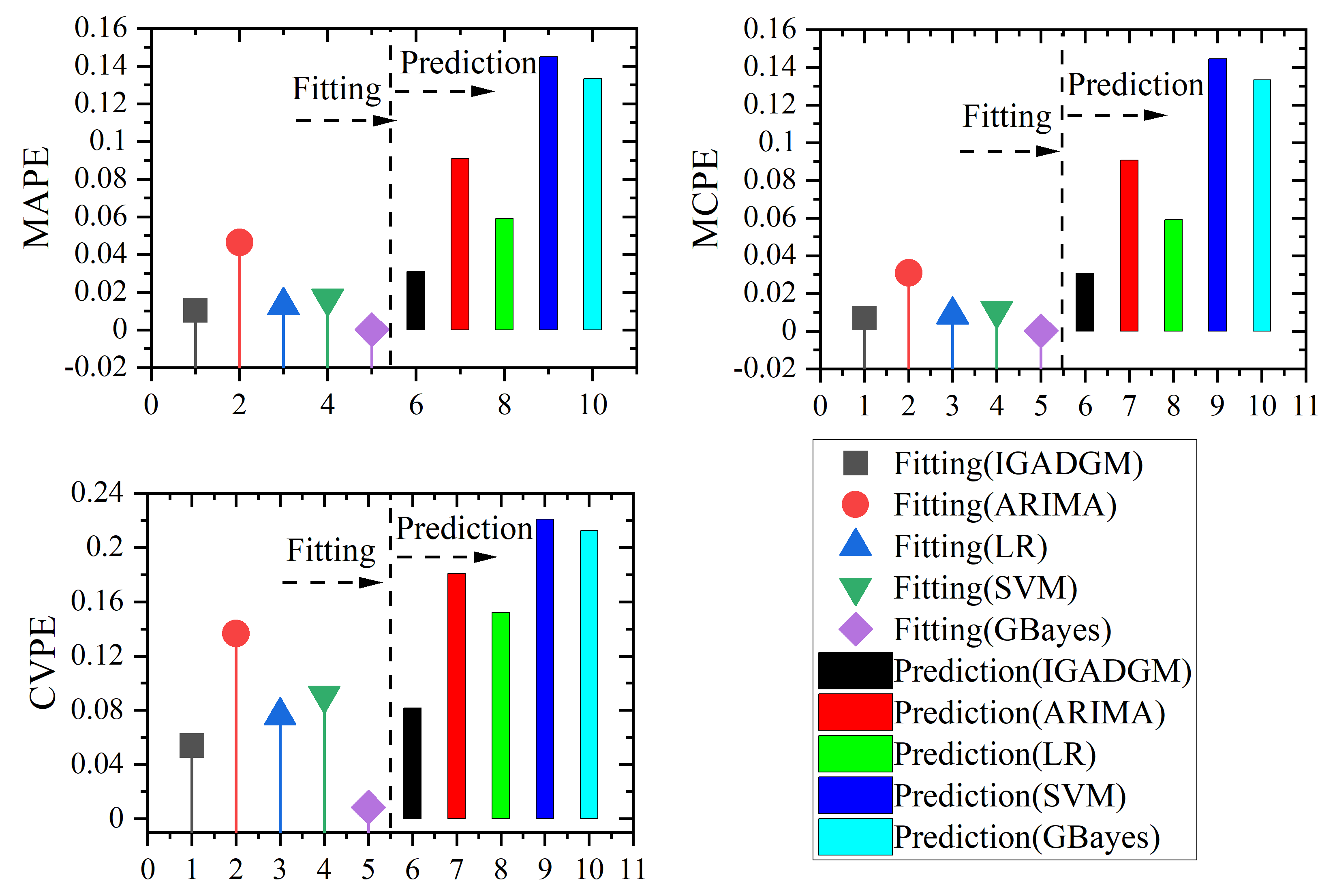} 
	\caption{The Error Evaluation of Three Models on India Crude Oil Import Structure} 
	\label{ar4}
\end{figure}

The contrast of five models and their error evaluation are shown in Table.\ref{chinaTable}, \ref{arcticError} in Figure.\ref{ar1}, \ref{ar2}, \ref{ar3}, \ref{ar4}, where the first six years are used to training the models and then the rest four years used for testing the models. It is clear to observe that the ARIMA, LR, GaussianNB, and GADGMSS model has good fitting precision but ordinary prediction precision. SVM model has general fitting and prediction precision. Despite an average fitting, IGADGM works way better than other four models with respect to prediction precision (less than 5\% among all of three error evaluations). The reason for better performance of IGADGM may derive from the fact that grey model indeed owns a better mechanism on short-term predicting. Moreover, through infusing the GADGMSS with TGMI, the novel model is be of good robustness and well predicted.  Therefore, it is reasonable to use IGADGM model to predict the proportion of crude oil import in China in the future.

\begin{table}[H]
	\centering
	\caption{Three Types of Error Evaluating Five Models Performance from Fitting and Prediction Respectively}
	\label{arcticError}
	\setlength{\tabcolsep}{7mm}{
	\begin{tabular}{ccccccc}
		\toprule
		& \multicolumn{3}{c|}{Prediction Error}            & \multicolumn{3}{c}{Fitting Error}                \\ \cline{2-7} 
		Models     & CVPE           & MCPE           & MAPE           & CVPE           & MCPE           & MAPE           \\ \hline
		ARIMA       & 0.091          & 0.091          & 0.180          & 0.046          & 0.031          & 0.137 \\
		IGADGM     & \textbf{0.031} & \textbf{0.031} & \textbf{0.082} & 0.010          & 0.007          & 0.054          \\
		LR       & 0.059          & 0.059          & 0.152          & 0.013          & 0.009          & 0.076          \\
		SVM        & 0.145          & 0.144          & 0.221          & 0.017          & 0.011          & 0.091          \\
		GaussianNB & 0.133          & 0.133          & 0.212          & \textbf{0.000} & \textbf{0.000} & \textbf{0.008}          \\ \bottomrule
	\end{tabular}
}
\end{table}

From Figure.\ref{ar4}, compared to the other four popular predicting models, the IGADGM reaches the smallest error with regards to three metrics and have a better performance on predicting China's crude oil import structure. The accuracy of IGADGM is nearly twice as the model with second smallest model, which as well proves the validation of IGADGM. In IGADGM, the \emph{DF}=5.998 means the weight of results from GADGMSS in IGADGM accounts for nearly 60\%, excceding to that from TGMI, which indicates that the China's crude oil import structure is being changed into a closely interconnected trend. Furthermore, development parameter (0.521) provides the trend with a quantitative speed, the overall change rate for the complete system is 0.521 year-on-year (the system would be stable if the rate is around 1).

As Table.\ref{arcticprediction} and Figure.\ref{ar3} shows, IGADGM is used to predict the structure of crude oil prediction in the next 6 years. The main source of import is still the Middle East, but the proportion of crude oil import from the Middle East and Africa will gradually decline. Meanwhile, the proportion of crude oil import from the arctic will maintain a relatively stable growth. According to the result,  reasonable opinions and suggestions are put forward on imported crude oil to reduce the transportation cost of imported crude oil and save financial expenses.

\begin{table}[H]
	\centering
	\caption{The  Prediction on Composition Structure of China's Import Crude Oil under IGADGM (2020 to 2025)}
	\label{arcticprediction}
	\setlength{\tabcolsep}{7mm}{
	\begin{tabular}{cccccc}
		\toprule
		Years & Arctic field & Middle East & Africa & Americas & Southeast Asia \\ \hline
		2020  & 0.195        & 0.454        & 0.117  & 0.116    & 0.118          \\
		2021  & 0.196        & 0.442        & 0.121  & 0.118    & 0.123          \\
		2022  & 0.202        & 0.436        & 0.117  & 0.121    & 0.124          \\
		2023  & 0.210        & 0.430        & 0.113  & 0.123    & 0.124          \\
		2024  & 0.217        & 0.424        & 0.109  & 0.125    & 0.125          \\
		2025  & 0.225        & 0.418        & 0.105  & 0.127    & 0.125       \\  \bottomrule
	\end{tabular}
}
\end{table}

\subsubsection{Empirical Analysis}

\begin{table}[!htb]
	\centering
	\caption{The Current Arctic Import Routes of Crude Oil and Their Miles} \label{routes}
	\resizebox{\textwidth}{!}{
		\begin{tabular}{ccc} 
			\toprule
			Name&Route&Voyage(sea mile)\\
			\midrule
			Middle East route 1& Kuwait port-strait of Hormuz-Singapore-Lianyungang&6371\\
			Middle East route 2&Suez-Mander strait-Singapore-Lianyungang&7362\\
			West Africa&Lagos-Cape Town-Singapore-Lianyungang&10588\\
			The South America&the port of Maracaibo-panama canal-Paific-Lianyungang&9527\\
			The Arctic passage 1&Murmansk NSR-Bering strait-Lianyungang&7355.4\\
			The Arctic passage 2&NSR Rotterdam-Bering strait-Lianyungang&8992\\
			\bottomrule
			
	\end{tabular}}
\end{table}

\begin{enumerate}
	\item Arctic Increases in Crude Oil Imports
    \begin{itemize}
    	\item \textbf{Abundant Crude Oil Reserve} In 2000, the US geological survey (USGS) estimates that 23.9\% of the world's undiscovered
    oil and gas resources were buried in the Arctic. Followed by May 2008, a survey conducted by U.S. shows the Arctic may contain 90 billion barrels of undiscovered oil, thus accounting for about 13\% of the world's undiscovered oil total today.
    Although the development progress in the Arctic is very conservative, much more focus should  nevertheless be put on the potential of making arctic passage possible down the road. 
    	\item \textbf{Shortening Transportation Distance}
    In the Table. \ref{routes}, compared with the traditional route via the Suez, the distance from Murmansk via the northeast passage to Shanghai can be shortened by 40\%. With the distance from Shanghai to Lianyungang ignored, the total distance over the new arctic passage is approximately 7,355.4 nautical miles while the traditional route distance of 12,259 nautical miles. Moreover, the Arctic shipping routes can also save the distance from Europe to China's coastal ports, among which the distance from Rotterdam to Lianyungang via the northeast shipping route is about 8,992 nautical miles. That's nearly 1,750 nautical miles shorter than the traditional route.
    \end{itemize}
    
    \begin{figure}[H]
    \centering
    \includegraphics[width=.6\textwidth]{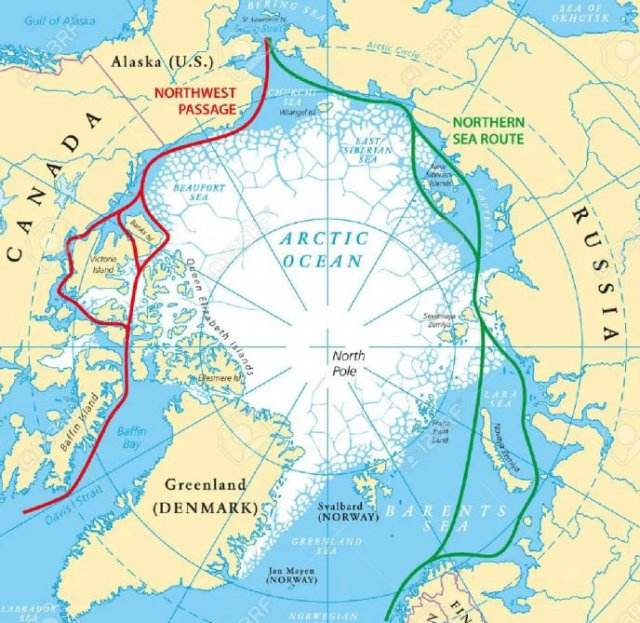}
    \caption{The Two Arctic Routes: Northwest Passage \& Northeast Passage}
    \end{figure}
    
    \item Declines in Crude Oil Imports from the Middle East
    \begin{itemize}
        \item \textbf{Risky Transportation Routes} Since nearly 70\% of China's crude oil imports come from middle eastern and African countries, the bulk of crude oil transport is also concentrated on routes from those regions to China. In addition, terrorist attacks and piracy are another prominent risk of maritime transport. In 2009, there were 406 attacks, half of them happened at the coast of Somalia, where crude oil is highly risky to transport by sea\cite{Middleeast}. 
        \item \textbf{"Asian Premium" on Imports} As the main importer of crude oil in the Asia-Pacific region, China has been suffering from the erosion of "Asian premium" in the long-term crude oil trade,\cite{Premiums} resulting in a large amount of capital outflow. In 2017, China imported $6.41\times10^{7}t$ crude oil from Saudi Arabia. According to this calculation, the loss of Saudi Arabia alone due to the "Asian premium" brought to China is more than 450 million dollars.\cite{asiapremiums}
    \end{itemize}

    \item Declines in Crude Oil Imports from Africa
    \begin{itemize}
        \item \textbf{Fully Tapped Potential} In 2018, China's crude oil imports from Africa were only 43.55\% of those from the gulf states.  More than two-thirds of Algeria's oil and gas resources remain under-explored or unexplored\cite{structure}. The imports of crude oil from Africa will slow down due to insufficient exploitation.
        \item \textbf{Unstable Political Situation} Among the 15 countries from which China imports crude oil in Africa, except South Africa, other countries have experienced long-term political instability, political party disputes, internal conflicts, ethnic disputes and international and regional tension. As a result, the security situation was worrying, and the scale of crude oil exploitation and export was sharply reduced, which directly affected the security of China's crude oil import from Africa \cite{doingbusiness}. 
    \end{itemize}
    
\end{enumerate}

\section{Conclusion and Future Work} \label{sec4}

Aimed to extend GM(1,1) model into Simplex space with Aitchison geometry, the discrete form of GM(1,1) is introduced, and as a result of this paper, GADGMSS is proposed. Furthermore, the infusion of GADGMSS and TGMI is adopted in order to improve the model's interpretability before its application of real datasets. In a nutshell, the main contributions of this paper are briefly summarized as follows:

\begin{enumerate}
    \item The concept of compositional data, a type of grey data, introduced into energy structural analysis. In comparison to directly predicting the white data, the components of different region can not only reflect their proportion in an overall, but interaction among regions can be quantitative calculated.
	\item  From the perspective of compositional data-driven, IGADGM is proposed by combining TGMI and GADGMSS according to their error's weight, which ensures a good fitting of the model. Moreover, not only does IGADGM take advantage of the correlation of different dimensions, which is attested by $\hat{\beta}_1=(\mathbf{\dot{X}}^{(1)}_0,\mathbf{\dot{X}}^{(1)}_0)^{-1}_S(	\mathbf{\dot{X}}^{(1)}_0,	\mathbf{\dot{X}}^{(1)}_1)_S$ , but shares heterogeneity of \emph{n-1} sub-models.   
	\item  The optimal value of the general accumulated matrix is searched by employing DE algorithm, which shows an outstanding performance over optimizing \emph{O($n^2$)} coefficients. Moreover, the IGADGM has a better efficacy than GADGMSS, TGMI, ARIMA, SVM, GaussianNB, and LR, under the limited number of sample.
	\item China's crude oil import structure is predicted by IGADGM from 2020 to 2025 and it turns out that the proportion of crude oil import from the Arctic area will maintain a relatively stable growth. The mild growth may come from two points: Arctic ares' vast reserve of crude oil and shortening distance thanks to Arctic passage. Moreover, the IGADGM model can be applied for structural analysis in more application fields such as gas, mineral prediction and even the industrial structure down the road. 
\end{enumerate}

Despite many advantages for IGADGM, there is also some future work for IGADGM as its first arising in the grey system extension. The power GM(1,1) model and other advanced GM(1,1)'s models will be tried in Simplex space with Aitchison Geometry.

\section{Acknowledge}
The author would like to appreciate the support from geatpy\cite{geatpy}, which has provided us with DE algorithm's API. Moreover, this work is supported by National Natural Science Foundation of China under grant 71871174 and 61403288 and Grey Thematic Innovation Zone under grant GS2019006 (China).

\bibliography{sample}



\end{document}